\documentclass[11pt]{article}
\pdfoutput=1
\usepackage{jheppub}
\usepackage{multicol}
 \def\beq{\begin{equation}\begin{aligned}}
 \def\eeq{\end{aligned}\end{equation}}
 
\begin{document}
\title{Annihilation Signals from Asymmetric Dark Matter}

\author[1]{Edward Hardy,}  
\emailAdd{e.hardy12@physics.ox.ac.uk}
\author[1]{Robert Lasenby,}
\emailAdd{robert.lasenby@physics.ox.ac.uk}
\author[2]{and James Unwin}
\emailAdd{james.unwin@nd.edu}
\affiliation[1]{Rudolf Peierls Centre for Theoretical Physics,
University of Oxford,\\
1 Keble Road, Oxford,
OX1 3NP, UK}
\affiliation[2]{Department of Physics, 
University of Notre Dame,\\ 
225 Nieuwland Science Hall, 
Notre Dame, IN 46556, USA}




\abstract{
In the simplest models of asymmetric dark matter (ADM) annihilation
signals are not expected, since the DM is non-self-conjugate and the
relic density of anti-DM is negligible. We investigate a new class
of models in which a symmetric DM component, in the `low-mass' 1-10
GeV regime favoured for linking the DM and baryon asymmetries, is
repopulated through decays. We find that, in models without significant
velocity dependence of the annihilation cross section, observational
constraints generally force these decays to be (cosmologically) slow.
These late decays can give rise to gamma-ray signal morphologies
differing from usual annihilation profiles. A distinctive feature
of such models is that signals may be absent from dwarf spheroidal
galaxies.
}

\maketitle


\section{Introduction}
\label{sec:intro}

Asymmetric dark matter (ADM) provides a well motivated framework for
light DM and is an intriguing alternative to the usual WIMP scenario.
In such models the DM, which we denote here $B'$, has an (approximately)
conserved quantum number (which we also call $B'$). The relic density is
determined by a particle-antiparticle asymmetry between $B'$ and $\overline
{B'}$, in direct analogy to baryons \cite{review}. If the DM has a similar
mass to the proton $m_{B'}\sim m_{p}$ and the hidden and visible sectors are
connected via portal operators which violate $B$, $L$ and $B'$, but
conserve some linear combination, then this can explain the cosmological
coincidence $\Omega_{\rm DM}\sim5\Omega_{B}$.  
In contrast, the accidental proximity of $\Omega_{\rm DM}$ and
$\Omega_{B}$ in conventional DM scenarios seems unreasonable given that
the relic density of DM is determined by freeze-out, whereas the baryon
density is set by CP-violating decays of out-of-equilibrium states and
these two mechanisms are typically unrelated.

While we shall not specify the UV physics that produces
these particle asymmetries, there are several 
mechanisms that can {\em cogenerate}
equal magnitudes for $\eta_{B'}$, $\eta_{B}$ and/or $\eta_{L}$ 
(see e.g.~\cite{cogenesis}). Alternatively, an asymmetry could be
generated in a single quantum number and subsequently shared via
processes that violate the individual global symmetries, leading
to comparable asymmetries. For the
case $\eta_{B'}\simeq\eta_{B}$, to account for the  DM relic 
density  one requires the DM mass to be around
5 GeV. Inefficient sharing, or bias generation, of the DM
and baryon asymmetries may easily lead to $\mathcal{O}(1)$ deviations between
$\eta_{B'}$ and $\eta_{B}$ and hence the DM mass could reasonably lie in
a relatively large window 0.1-100 GeV. However, arguably the most natural
mass region is 1-10 GeV.

Indirect detection signals of DM can arise if the DM decays or
annihilates producing cosmic rays containing high energy photons,
electrons, positrons or antibaryons. Usually, the event density associated to
signals of decaying DM depends linearly on the DM density 
$n_{\rm DM}$, whereas for annihilating DM the signal has an $n_{\rm
DM}^2$ dependence. If astrophysical gamma-ray signals from DM were detected,
then their profile on the sky would let us determine which process produced them, 
e.g.~\cite{Boehm:2010qt,Buchmuller:2012rc} (the propagation
of charged cosmic rays is affected by galactic magnetic fields, 
so a signal in these channels would not be so helpful).
Whilst decay signals of ADM can readily arise, see e.g.~\cite{Zhao:2014nsa}, 
in general, we do not expect late-time annihilation signals from ADM,
since the symmetric component is assumed to have annihilated early on.
However, if the symmetric component is later `regenerated',
this may give rise to annihilation signals. Previous proposals along
these lines include slow DM-antiDM oscillations~\cite{Oscillating},
and intermediate-time decays (in the higher DM-mass regime,
\cite{Falkowski:2011xh}). Alternatively, there are scenarios in which
an asymmetry plays a role in determining the DM relic density, but
the DM itself is not asymmetric~\cite{alt}. In this paper, we investigate the prospect of regenerating a
symmetric component via decays in the low DM-mass regime.

A challenge in building viable models of ADM is obtaining a high enough
annihilation cross section so that the relic density is set by the
asymmetry and not by a frozen-out symmetric DM component. For this to
occur the cosmologically stable states in the dark sector typically
must have annihilation cross sections some factor larger than
the thermal value ($\sim 3 \times 10^{-26} {\rm cm}^3 {\rm s}^{-1}$).
Experimental limits on this scenario arise from a variety of sources. Direct
searches rule out large regions of parameter space where
the interaction with the visible sector is through heavy portal states
that can be parameterised as effective operators, see
e.g.~\cite{MarchRussell:2012hi,Buckley:2011kk,LEPdm,ColliderEW},\footnote{For 
contact operators with preferential coupling to specific
SM states, such as neutrinos~\cite{Shoemaker:2013tda}, muons, or taus, some
of the tension with experimental constraints can be alleviated.}
although viable models remain if the mediating states are light.
Perturbativity also often limits the cross sections that can be obtained
in viable models. Additionally, indirect detection constraints are
limiting in some regions of parameter space.

Generally, the strongest astrophysical constraints on the DM 
annihilation cross section (if the population is entirely symmetric)
are close to or below the lower cross-section bound from the ADM
relic annihilation constraint.
In particular, if a sizeable symmetric component is regenerated before
the recombination era, then the limits on the annihilation cross
section from CMB observations \cite{CMB} are typically only a small factor above
the thermal freeze-out value, for DM masses $\lesssim 20~{\rm GeV}$. Since the
CMB limits are mostly a function of energy injection, rather than the
particular annihilation products, and are not subject to uncertainties
regarding DM distribution, they leave very little parameter space for early decays, for annihilation
cross-sections large enough to meet the ADM relic annihilation requirements.

One manner of circumventing these constraints is for the
annihilation cross section at late times to be suppressed relative
to earlier times, e.g.\ via velocity-suppression.
If the symmetric population is regenerated after freeze-out,
 but early relative to astrophysical timescales, then the phenomenology of this scenario is
much like symmetric DM with a velocity-suppressed cross section, 
although with the annihilation rate decoupled from the
thermal value. An alternative possibility, with distinctive
phenomenology, is the case in which the DM remains predominantly asymmetric 
up to the present time (or at least through the recombination era), with only 
a small symmetric component regenerated.
If the decay timescale is significantly longer than the time at
recombination, then CMB constraints are significantly weakened, which
can open the possibility of present-day annihilation signals without
corresponding CMB signals.
Another interesting feature of regenerating the symmetric component through
late decays, which we will discuss in some detail, is that its density
profile may be modified significantly by the velocity kick from the
decay process. Obtaining a sufficiently small velocity kick to keep the
regenerated population concentrated in galaxies motivates DM models
with a small mass splitting between different components. For these models, the
annihilation profile may be different to those arising from conventional
models, and could be entirely absent from systems with low escape
velocities such as dwarf spheroidal galaxies.

As mentioned, coupling ADM to the Standard Model (SM) without
violating direct search constraints requires
some model-building, and we consider two scenarios in detail. In one case the
annihilation portal to the SM is through a pseudoscalar; 
this alleviates the direct detection constraints, as the DM-proton 
scattering cross section is significantly suppressed.
If the pseudoscalar mass is just above the value needed
for an s-channel annihilation resonance, then annihilations during the freeze-out
process (when temperatures are high) are enhanced relative to annihilations at
later times. 
This is essentially an example of the velocity dependence discussed above.
The second scenario we
consider is where annihilation occurs to two on-shell hidden sector vector bosons, which decay relatively slowly to SM states. The small coupling of the vectors to the SM suppresses direct search constraints, and the rate at which we regenerate the symmetric component controls the strength
of CMB and late-time annihilation constraints.
These models act as a {\em proof
of principle} that annihilation signals can be generated from certain
models of ADM, but the constraints that we discuss are more generally relevant to other models of annihilating DM in the low mass region. 
The observation of signals in this mass region may suggest that the DM
relic density is set by an asymmetry, and that the hidden sector dynamics is correspondingly 
more complicated than often assumed.

Turning to the structure of the paper, in Sect.~\ref{S1} we begin by
discussing an example class of models that regenerates a symmetric
component through decays. Subsequently, in Sect.~\ref{S2} we explore the
constraints on models of annihilating ADM and identify the parameter
space for which successful models can be constructed. Additionally,
we study the modification of signal morphologies due to the decays.
Sect.~\ref{S3} explores the hidden sector model-building necessary to
satisfy the constraints obtained.


\section{Regeneration from models with two dark asymmetric species}
\label{S1}

To construct a model with late time decays, regenerating a symmetric
component of DM, requires a more complicated hidden sector. By
simple analogy to the complexity of the visible sector it is quite
conceivable that the hidden sector consists of multiple states and
approximately conserved global currents. As a model building example
we consider a particular setup, but the constraints and possible
signals we find in Sect.~\ref{S2} are more generally applicable. 

Let us suppose that two
states $B^\prime,~L^\prime$ in the hidden sector carry (approximately)
conserved quantum numbers which we suggestively call dark baryon
number $B^\prime$ and dark lepton number $L^\prime$.  In the SM, baryon and lepton number are approximately conserved; EW instantons and sphalerons, and GUT scale physics do not respect these global symmetries, however these processes still conserve the combination $B-L$. In direct analogy, we propose that
both $B^\prime$ and $L^\prime$ are accidental symmetries of the low energy theory and are violated at some intermediate scale,
however the combination $B^\prime-L^\prime$ is conserved by these
intermediate scale interactions. We further assume that there are additional  effects in the theory that link the asymmetries in the dark and visible sector, so that the true symmetry of the theory is $B-L-(B'-L')$. Since it is expected that all global symmetries are violated by $M_{\rm Pl}$-effects, the combination  $B-L-(B'-L')$ may either be gauged in the UV or violated only by  $M_{\rm Pl}$ effects that are not important for phenomenology.

\begin{figure}[t!]
\begin{center}
\includegraphics[width=0.5\textwidth]{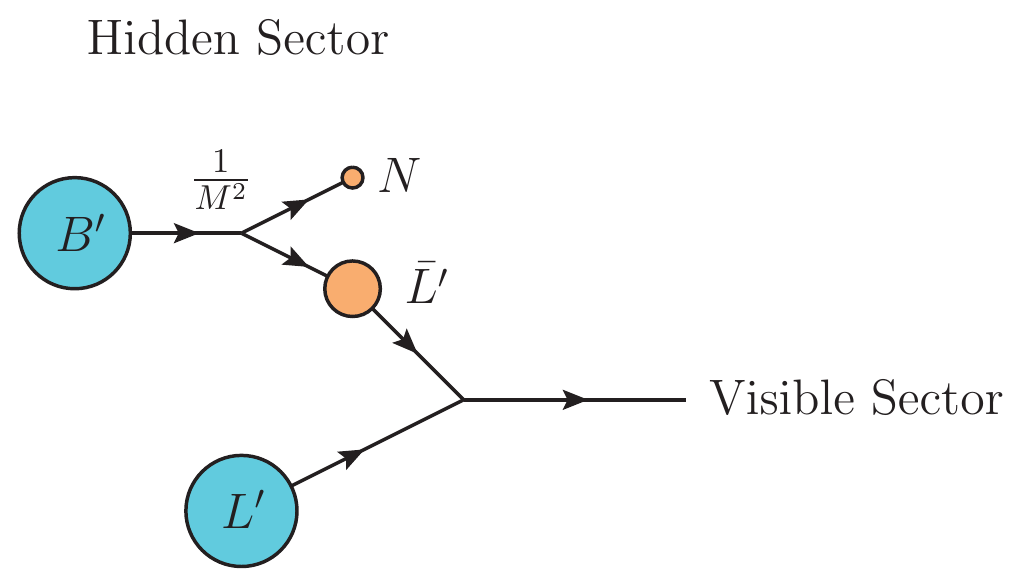}
\caption{Schematic illustration of steps leading to annihilation signal.
Initially, $B'$ and $L'$ are the only states in the hidden sector with
non-negligible abundances. 
 At some later time, such that it occurs in
the current era, $B'$ decays to $\overline{L'}$ along with additional light states labelled $N$. The $\overline{L'}$ subsequently (on
galactic time scales) annihilate with $L'$ to visible sector states
leading to potentially observable signals.}
\label{Fig1b}
\end{center}
\end{figure}

Various potential genesis mechanisms have been outlined in the literature,
which could be employed to generate the particle asymmetries e.g.~\cite{ADMgenesis}. 
We shall suppose that the asymmetries in $B'$ and $L'$ are of comparable magnitude,
and that there is no overall asymmetry in the quantity $B-L-(B'-L')$. 
For sufficiently high annihilation cross-sections, the $B'$ and $L'$ states 
remain chemically coupled to the thermal bath long enough for Boltzmann
suppression to remove most of their symmetric components.
This results in $\overline{B'}$ and $\overline{L'}$ having highly suppressed
relic abundances, whilst the yields of $B'$ and $L'$ are set by the particle-antiparticle asymmetries
\beq
\frac{n_{B'}}{n_{\gamma}}\simeq\eta_{B'}~, \qquad \frac{n_{L'}}{n_{\gamma}}\simeq\eta_{L'}~.
\eeq 
We will also assume that $\eta_{B'} \simeq \eta_{L'}$, so that we
end up with a relic population of $B'$ and $L'$ rather than their
anti-particles. Asymmetries with different signs and magnitudes are also
possible (and may be brought about dynamically, as discussed below), but
we will not consider such models in this paper.

For such a scenario to lead to annihilation signals we propose that $B'$-violating operators, suppressed by some intermediate scale, induce the decay of $B'$ to $\overline{L'}$, which subsequently annihilate with the population of $L'$ resulting in observable signals. This scenario is illustrated schematically in Fig.~\ref{Fig1b}. This is inspired by proton decay due to GUT scale physics \cite{Nath:2006ut}, from exchange of an $X,Y$ boson, inducing a decay $p\rightarrow e^++\pi^0$ which is $B$ and $L$ violating but $B-L$ conserving. The proton lifetime is thus dependent on the GUT scale, the scale at which $B$ is perturbatively violated
\beq
\tau_p\sim\frac{M_{\rm GUT}^4}{\alpha_{\rm GUT}^2 m_p^5}~.
\label{gut}
\eeq
We expect a similar expression for the lifetime of $B'$ decaying to
$\overline{L'}$, dependent on the scale of $B'$ violation (to some
appropriate power, set by the dimension of the decay operator).

The case of particular interest is when the decays $B'\rightarrow
\overline{L'} + \cdots$ (where the ellipsis denotes additional,
relatively light, decay products)
are slow such that this process occurs after the (dark) matter has
coalesced into galaxies and clusters. If the decay products acquire
too much kinetic energy, they will travel faster than
galactic escape velocity. 
In this case, as discussed later, they are unlikely to annihilate on their way out
of the galaxy, resulting in either decay-type profiles or no observable signals.
The maximum kinetic energy available 
is determined by the mass splitting $\Delta m= m_{B'} - m_{L'}$, and a natural
way to ensure that the final velocities are low enough, so to enable
observable annihilation signals, is to have $\Delta m$ small relative to the DM mass.
Such small mass splittings can arise
for instance through radiative mass splitting between different DM `flavours'. 
 We discuss this constraint more generally in the next section, and outline two models
which exhibit small mass splittings in Sect.~\ref{S3}. 
Additionally, the small mass splitting will suppress the $B'$ decay
rate. For example, if the decay is via a
dimension-6 operator induced by $B'$-violating physics at the scale
$M_{\not B'}$, the lifetime will be (parametrically)
\begin{equation}
\tau_{B'} \sim 4 \pi \frac{M_{\not B'}^4}{\alpha^2 (\Delta m)^5}~,
\label{lifetime}
\end{equation}
where $\alpha$ is the coupling associated to the $B'$-violating physics.

The fact that the $B'$ decay rate is suppressed (by powers of $\Delta
m$) relative to symmetry-violating processes at higher energies means
that it is necessary to check that the processes in the early universe
do not alter the $B'$ and $L'$ asymmetries in undesirable ways.
Indeed, as discussed in Appendix~\ref{app:rates}, if these interactions are in equilibrium after
the asymmetries have been established, they force $\eta_{B'} = -
\eta_{L'}$, rather than the same-sign asymmetries required for the
models discussed here. Consequently, the asymmetry must be set
sufficiently late, such that these symmetry-violating processes
are always out-of-equilibrium. Additionally, we assume that these
asymmetries are set before freeze-out,\footnote{The alternative case
of the asymmetry being set by decays after freeze-out is potentially
interesting, however we will not examine such models of here.} leading
to an upper limit on $\Gamma$ (equivalently, a lower limit on $\Delta
m$). We shall examine these constraints shortly and demonstrate that
viable models can be constructed.


\begin{figure}
\centering
\includegraphics[width=0.7\textwidth]{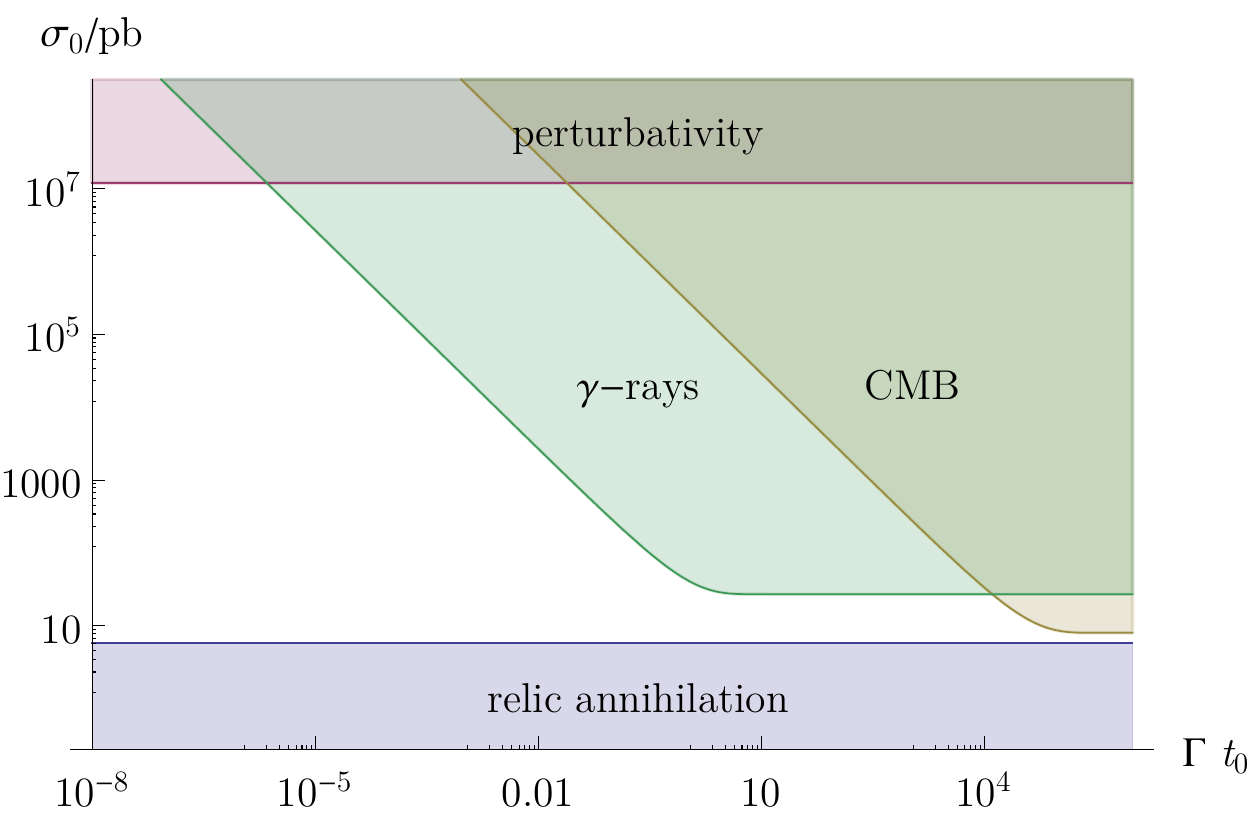}
\caption{Constraints on $B' \rightarrow \overline{L'}+\cdots$ decay rate
$\Gamma$ and $L' \overline{L'}$ annihilation cross section $\sigma_0$, with $m_{L'} = 10~{\rm GeV}$, assuming annihilation to muons. The
perturbativity bound corresponds to the model of Sect.~\ref{sec:Lpp},
where $L'\overline{L'}$ annihilate into a pair of hidden sector vector
particles, which then decay to the SM. Here, $t_0$ is the current age of the universe.
The relic annihilation bound corresponds to the annihilation cross-section needed
to annihilate all but 1 percent of the symmetric component, i.e.\ to obtain
$r_\infty \equiv \frac{n_{\overline{B'}}}{n_{B'}} < 0.01$.
The CMB bound corresponds to the limit on energy injection from $L' \overline{L'}$
annihilation derived from CMB perturbations~\cite{CMB},
while the gamma-ray bound is derived from FERMI observations~\cite{FERMIdiffuse}
(assuming an NFW profile for the DM halo). We assume that the $\overline{L'}$
injection velocity is small enough that the $\overline{L'}$
distribution is similar to the $L'$ one for the purposes of the gamma-ray
constraints; we use the FERMI diffuse observations rather than the
galactic centre observations, which would be less robust to profile modifications
(see Sect.~\ref{sec:splitting}).}
\label{fig:regions}
\end{figure}


\section{General constraints and signals}
\label{S2}

Having introduced the general structure of the models considered,
we now examine the relevant constraints and highlight particular features required
for observable annihilation signals to arise in such models. For a given $B'
\rightarrow \overline{L'} + \cdots$ decay rate $\Gamma$, the $\overline{L'}$ population
builds up (essentially) linearly over time until a significant fraction
of the $B'$ have decayed (unless the annihilation cross section is extremely high,
only a small proportion of the $\overline{L'}$ population that is regenerated ever annihilates).
This $\overline{L'}$ population translates into an upper
bound on the $L' \overline{L'}$ annihilation cross section from astrophysical
indirect detection signals (both late-time and recombination-time).
A lower bound on the annihilation cross section comes
from the requirement that the relic symmetric component is sufficiently 
depleted due to annihilations in the early universe. 
Further, demanding perturbative couplings places an additional upper bound on the cross sections.

These constraints fit together as illustrated in Fig.~\ref{fig:regions},
which shows the allowed regions in terms of the (velocity independent)
annihilation cross-section $\langle \sigma v \rangle \equiv \sigma_0$, and the $B' \rightarrow
\overline{L'} + \cdots$ decay rate $\Gamma$. The indirect detection constraints
will depend on the DM mass (which sets the number density, and energy
injection from annihilation), and on the annihilation channel (here,
we show annihilation to muons --- alternative channels will generally
improve the CMB bounds by a factor of less than 3, and the gamma-ray
bounds by less than a factor 10).
The perturbativity bound corresponds to a particular choice of model
(see Sect.~\ref{sec:Lpp}), while the relic annihilation
bound is weakly dependent on the DM mass.
We observe that there is a significant amount of parameter space in which viable models can implemented. The potential signal region is the area close to the current gamma-ray limits,
which can be probed by current and future experiments.
 Also, it can be
seen that relatively slow decays are required to evade CMB limits.
If $\Gamma t_0 \gtrsim \frac{t_0}{t_{\rm CMB}} \sim 5\times 10^{-5}$
(where $t_0$ is the current age of the universe, and $t_{\rm CMB}$ is the age at the recombination era),
corresponding to the region at the far right of the diagram, then
there is only a very small region of allowed parameter space.
Current CMB observations are far enough above the cosmic variance limit~\cite{CMB} that
future experiments should be able to close this gap for most decay channels.

A further possible constraint arises from DM self-scattering bounds
(see e.g.~\cite{Tulin:2013teo,SelfScatteringHalo}). The relationship between the
annihilation cross section and the self-scattering cross-section is
model-dependent, so may or may not introduce constraints within the
perturbative region. In the model with a light hidden sector mediator
considered in Sect.~\ref{sec:Lpp}, the scattering cross section can be
sufficiently large to introduce limits. If
we drop the assumption of perturbativity, then since the self-scattering
cross section will generally be comparable to or larger than the
annihilation cross-section, we obtain an upper bound on $\sigma_0$
of at most around $0.04~{\rm barn} \times \frac{m_{\rm DM}}{10~{\rm GeV}}$.
This is low enough that, if the $\overline{L'}$ is ejected from the galaxy,
only a small fraction of it will annihilate on the way out --- consequently,
the spatial distribution of annihilations will be approximately $\propto n_{L'}$,
so will resemble a decay profile~\footnote{Such models in which annihilations produce 
decay-type gamma-ray profiles may be interesting in themselves, though we have not investigated them
in any detail.}. Thus, large annihilation cross sections
do not remove the need for a small velocity kick if we want to generate
an annihilation-type gamma-ray profile.

As we will discuss in Sect.~\ref{S2.1}, some model-building is required
to make sure that direct constraints (from collider and direct-detection
experiments) do not place strong upper bounds on $\sigma_0$. For DM-SM
interactions via contact operators, these bounds are generally (for $m_{\rm DM}
\lesssim 30~{\rm GeV}$) below the lower bound from relic annihilation,
so would leave no allowed parameter space. Fig.~\ref{fig:regions} can
be thought of as corresponding to the model of Sect.~\ref{sec:Lpp},
where $L' \overline{L'}$ annihilate to a pair of hidden sector particles,
which then decay to the SM.  In this case, direct constraints
can be completely absent.


\subsection{$B'$-$L'$ mass splitting and $\overline{L'}$ distribution}
\label{sec:splitting}

Unless the velocity of the decay-produced $\overline{L'}$
is low enough, these states will be ejected from the galaxy, and not
sufficiently concentrated to give detectable annihilation-type signals.
The escape velocity for the Milky Way is $\sim 500~{\rm km} \, {\rm s}^{-1}$,
so the fractional $B'$-$L'$ mass splitting (assuming two-body decay, with the other product
having much smaller mass) should satisfy
\beq
\frac{\Delta m}{m_{L'}}~\lesssim~ \frac{500~{\rm km} \, {\rm s}^{-1}}{c} ~\approx~  2 \times 10^{-3}~.
\label{3.1}
\eeq
For mass splittings much smaller than this, the $B'$-to-$\overline{L'}$
velocity change will be small compared to the velocity dispersion of the
$B'$, so the distribution of the $\overline{L'}$ will be close to that
of the $B'$, and annihilation signals will have the standard $\propto
n_{B'}^2$ profile. However, for intermediate mass splittings, the extra
velocity will be significant, and will result in an altered distribution
for the $\overline{L'}$.~\footnote{As relaxation
times for galaxies such as the Milky Way are large enough that stellar
encounters are generally unimportant~\cite{BT}, the extra energy will
not be removed by gravitational interactions. Also,
the bounds on 
DM self-scattering~\cite{Tulin:2013teo,SelfScatteringHalo} mean that
a given particle can interact at most a few times over galactic
timescales, so it is only at the upper end of the allowed
scattering cross section region that appreciable energy is lost
this way. Such large self-scattering cross-sections also have
consequences for the dominant component of the DM distribution (pushing the halo
towards isothermality), and we will not consider the details of such
models here.} 
Qualitatively, the
distribution will be `puffed out', with clumpiness being smoothed out
and, moreover, systems with low escape velocities (e.g.~dwarf galaxies)
will not develop a bound population of $\overline{L'}$ states.

Going beyond the two-body decay case, we will obtain a distribution of
$\overline{L'}$ velocity kicks, so some fraction of the $\overline{L'}$
will be given higher-than-escape velocities. However, in general it is
still the case that, as long as the other final or intermediate states
do not have masses close to $\Delta m$, very few of the $\overline{L'}$
end up with significantly smaller velocities than $\Delta m / m_{L'}$.
For example, if $\Delta m = 0.1 m_{L'}$, then the fraction of the
$\overline{L'}$ getting velocity kicks of less than $2\times 10^{-3}$ is
generally smaller than $5\times 10^{-6}$
(see Appendix~\ref{ApA} for details).
So, in the absence of other `coincidences' assuring a small velocity kick,
a small mass splitting is needed
to obtain significant $\overline{L'}$ bound populations.
As discussed in Appendix~\ref{ApA}, instead of tuning the $B'$ and $L'$ masses
to be close to each other, we could also tune the masses of the other decay products
or intermediate states; however, a small mass splitting is more natural in 
many model-building contexts.

\begin{figure}
\begin{center}
\includegraphics[width=0.7\textwidth]{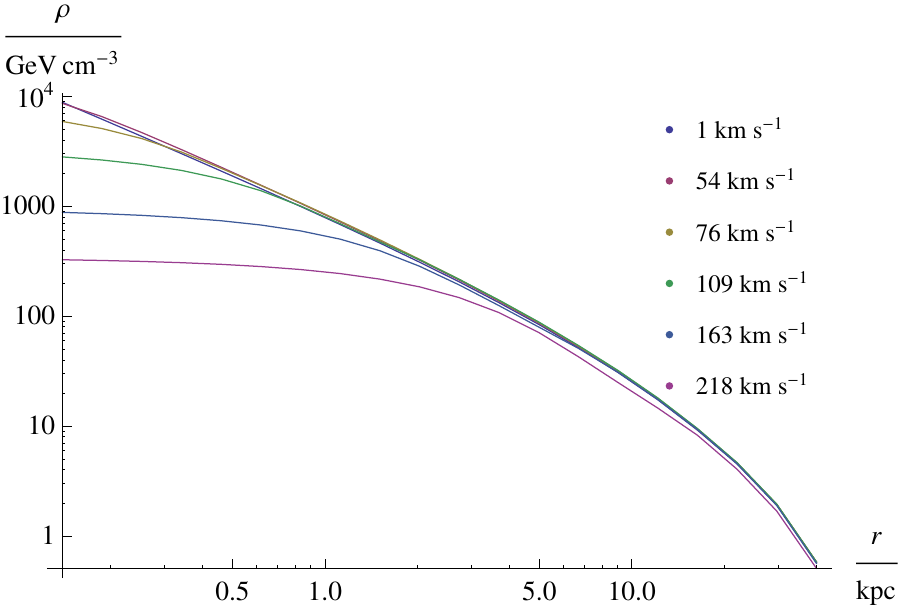}
\caption{Approximation to $\overline{L'}$ profile after injection at relative velocity $v_i$ as listed,
starting from a NFW-type $B'$ distribution with shape parameter $\gamma = 1.2$ (so 
$\rho \propto r^{-1.2}(1+r/r_s)^{-1.8}$, taking $r_s = 20~{\rm kpc}$).
The $B'$ distribution is taken to have isotropic velocity distribution, which
implies velocity dispersion of $\sigma = 94~{\rm km}~{\rm s}^{-1}$ at $r = 1~{\rm kpc}$.
As described in Appendix \ref{ApB}, we approximate the $\overline{L'}$ distribution as also
having isotropic velocity distribution, which will make the larger-$v_i$ profiles
slightly less peaked than they should be.
}
\label{fig:nfwAlpha}
\end{center}
\end{figure}

Quantitatively, Fig.~\ref{fig:nfwAlpha} shows an approximation to the
$\overline{L'}$ profiles obtained for different injection velocities,
starting from an NFW-type profile. As described in Appendix~\ref{ApB},
we calculate these by convolving the initial DM phase-space
distribution function with a velocity-kick kernel, then reparameterising
to find the new steady-state distribution function, and integrating
over this to find the new spatial number density. We see, as expected,
that the deviation from the distribution of the parent particle is
small for injection velocity smaller than the $B'$ velocity dispersion,
with the profile being flattened out for larger $v_i$. For the
example in Fig.~\ref{fig:nfwAlpha} the profiles of states produced
with $v\lesssim50~{\rm km~s}^{-1}$ remain approximately NFW and for
increasing injection energies the profiles are smooth deformations
away from the profile of the parent particle. In Sect.~\ref{3.6}, we
study the effect of this on the observed annihilation profile and
the compatibility with tentative signals that may have recently been
observed.


\begin{figure}
\begin{center}
\includegraphics[width=0.70\textwidth]{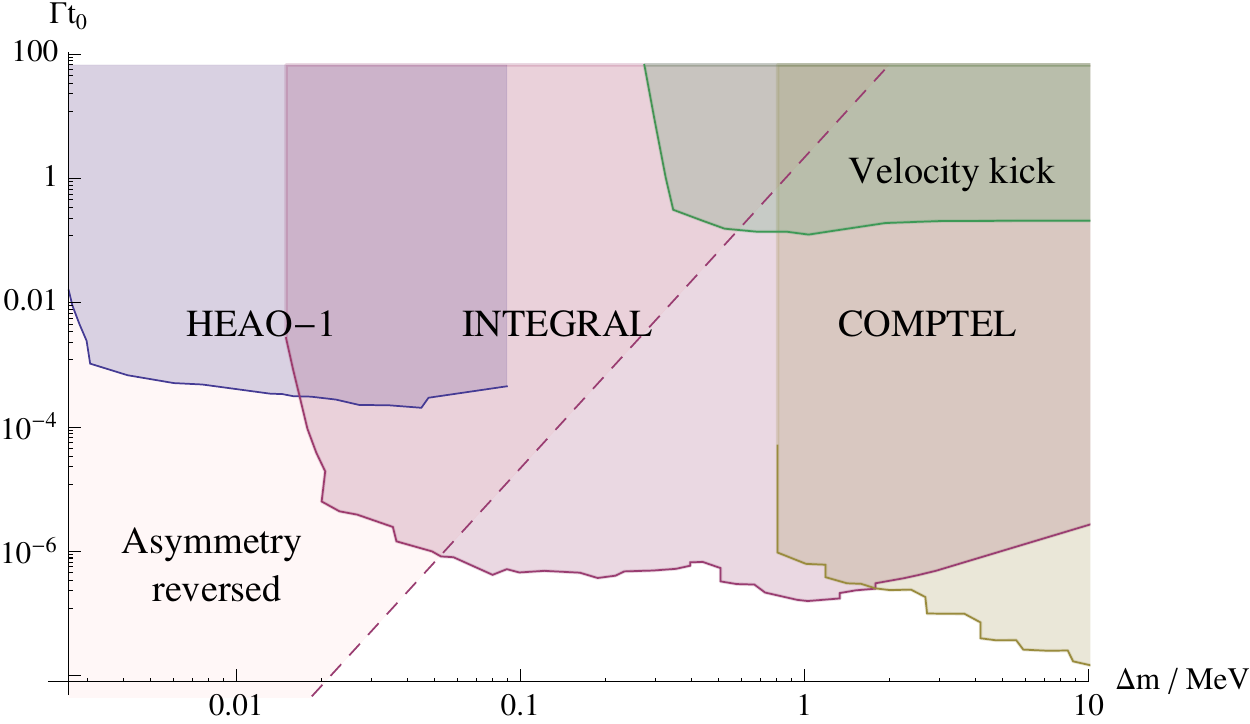}
\caption{Constraints on the $B' \rightarrow \overline{L'} + \cdots$ decay rate
$\Gamma$ (compared to the age of the universe $t_0$) from X-ray
observations, for $m_{B'} = 10~{\rm GeV}$, assuming that all decays are $B' \rightarrow \overline{L'} +
\gamma\gamma$ and that the DM profile is NFW; constraints from the HEAO-1 (blue),
INTEGRAL (red), and COMPTEL (yellow) experiments~\cite{Essig:2013goa} are shown.
The green region corresponds to velocity kicks (assuming $v_i \approx \Delta m /m_{B'}$)
ruled out by structure formation constraints~\cite{PeterSatellites,WZ1}.
The region above the dotted line is not viable in the model of Sect.~\ref{S3.1},
since $B'$ and $L'$ violating interactions stay in equilibrium until $T \lesssim 50\,{\rm GeV}$
(see Appendix~\ref{app:rates}).}
\label{fig:decSM}
\end{center}
\end{figure}


There is also the possibility of indirect signals from the $B'
\rightarrow \overline{L'}+\cdots$ decays themselves, if the other decay
products include SM states~\cite{ddmLifetime}. In Fig.~\ref{fig:decSM} we show conservative
(`worst-case') constraints on the decay rate in this scenario; we
assume that all decays proceed via $B' \rightarrow \overline{L'} +
\gamma\gamma$, producing a sharp spectrum of photons and maximising
the detectability of the signal. Note that, since the bounds from
decays to electrons are at most a factor of $\sim 100$ worse than
from those to photons~\cite{ddmLifetime}, then in the absence of a small
($\lesssim0.1$ MeV) mass splitting, if annihilation signals are to
dominate over decays it is a requirement that most of the energy from
decays is dumped into other hidden sector states (or into neutrinos).
Otherwise, since only a small proportion of the emitted $\overline{L'}$
have annihilated by the present day, the SM states emitted in every decay
will be a stronger signal.
This can impose an
extra constraint on our model building from the requirement of having
additional light stable hidden sector states, and ensuring these have
sufficiently small relic abundance. Such light states can arise 
through small Dirac or Majorana masses, or
as pseudo-goldstone bosons.

If the $B'$ lifetime is not much longer than the age of the universe,
then sufficiently large velocity kicks from its decay can affect
the structure formation process. The strongest constraints on
small velocity kicks ($v_i \lesssim 100\,{\rm km}\,{\rm s}^{-1}$) 
comes from their effect on the population of Milky Way satellite
galaxies~\cite{PeterSatellites,WZ1}. Such constraints depend on
modelling of the highly non-linear regime of structure formation, but
constraints from Lyman-$\alpha$ observations (which probe much earlier
times) are almost as limiting~\cite{LyAlpha}. Fig.~\ref{fig:decSM} shows
the galactic bounds in the $(\Delta m,\Gamma)$ plane (assuming $v_k
\approx \Delta m / m_{B'}$).

As commented on in the previous section, the requirement that
the asymmetries in $B'$ and $L'$ are not disrupted through
symmetry-violating processes in the early universe places a lower
bound on $\Delta m$ (equivalently, upper bound on $\Gamma$). In
Fig.~\ref{fig:decSM} we also show the approximate bound from requiring
that symmetry-violating interactions decouple sufficiently early,
as discussed in Appendix~\ref{app:rates}. The mass splittings in
Fig.~\ref{fig:decSM} are related to the velocity kicks, e.g.\ as
studied in Fig.~\ref{fig:nfwAlpha} by eq.~(\ref{3.1}). For example,
taking $m_{B'}= 10$ GeV, the value used in the figure, $\Delta m\approx
2$ MeV (7.2 MeV) corresponds to a velocity kick in the region of
$v=54~{\rm km~s}^{-1}$ ($218~{\rm km~s}^{-1}$). We can see that there is
a significant allowed region for all cosmologically-slow decay rates and
mass splittings in the range of interest. However, in combination
with the velocity kick bounds described in the previous paragraph, cosmologically
early decays are generally excluded, independently from the previously mentioned direct
detection bounds.


\subsection{Direct and indirect detection}
\label{S2.1}

\begin{figure}
\begin{center}
\includegraphics[width=0.65\textwidth]{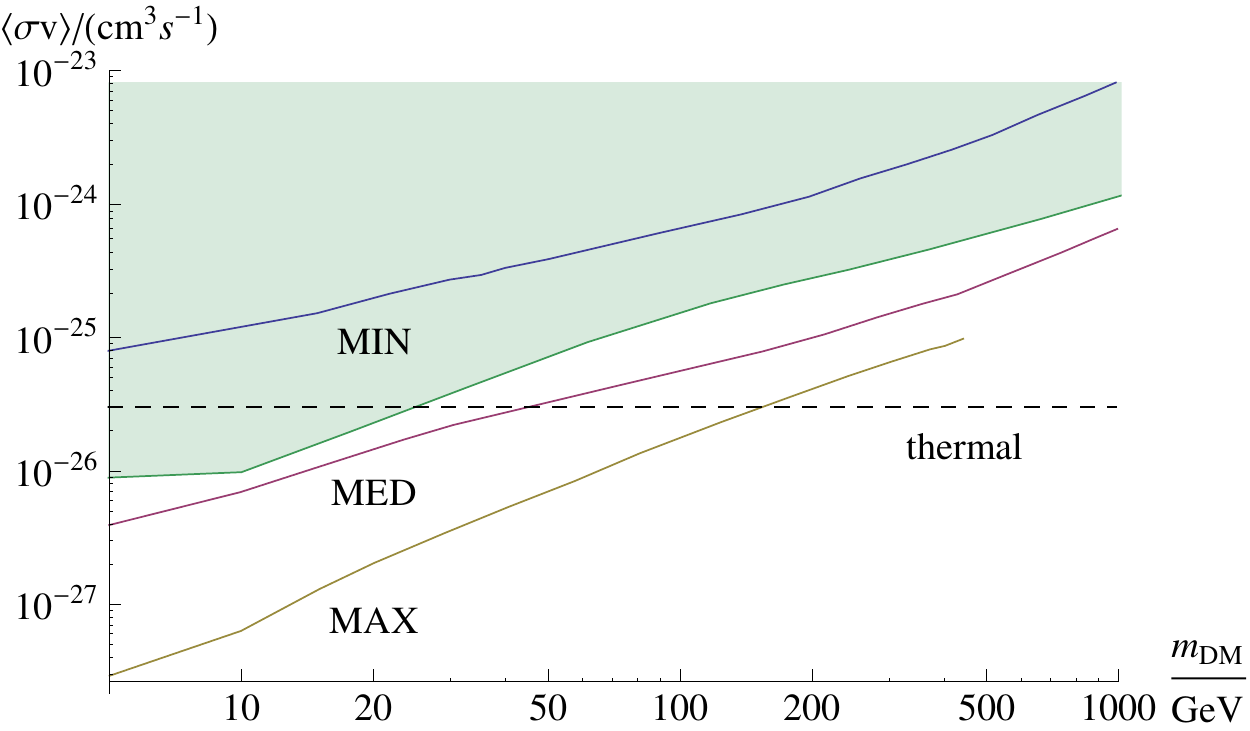}
\caption{Limits on symmetric DM annihilation cross section to $b\overline{b}$ from (blue, red, yellow) PAMELA cosmic ray antiproton
measurements~\cite{DMpbar}, under different assumptions with respect to propagation
of charged particles through the galaxy, and (green) FERMI gamma-ray
observations~\cite{FERMIdiffuse}. (Note that the PAMELA bounds correspond to assuming
that the DM halo has an Einasto density profile, while the FERMI bounds assume
a NFW profile; however, the difference will be minor~\cite{Evoli:2011id}.)}
\label{fig:pbar}
\end{center}
\end{figure}

In addition to the gamma ray constraints arising from late-time annihilations, there is also the prospect of signals coming from annihilation products in cosmic rays
(for the low DM mass range we consider, annihilations into neutrinos will
generally be beyond the reach of near-term experiments~\footnote{If the
DM is sufficiently strongly self-interacting, and has some scattering cross-section
with nuclei, there is the possibility
of building up a large population within the Sun, annihilations within which may
be detectable by neutrino observatories~\cite{Zentner:2009is,Albuquerque:2013xna}.
Up to some threshold, lower annihilation rates will actually increase this signal
(by allowing a larger equilibrium DM population in the Sun), so such signals could arise
for decay rates much slower than the age of the universe.}). DM annihilation into quarks results in cosmic ray antiprotons, the
population of which has been measured by the PAMELA experiment
(and will also be measured by AMS-02) \cite{DMpbar}. Since antiprotons are charged
particles, whose trajectories are affected by galactic magnetic fields,
their direction of arrival is not simply related to the location of
their source, and very little can be inferred about the galactic
DM distribution from such measurements. In addition, there is a
considerable degree of uncertainty as to exactly how this propagation
through the galaxy occurs, and different models result in
significantly different derived constraints on DM annihilation rates~\cite{Evoli:2011id}.
Fig.~\ref{fig:pbar} illustrates how the antiproton limits compare to
those from gamma-ray observations, under three different propagation
scenarios (MIN/MED/MAX) that are standard in the literature, see e.g.~\cite{DMpbar}. The limits range from significantly less constraining than those from gamma-rays, to very significantly more so. 
Consequently, depending on the true propagation, it is possible that annihilation signals may first be observed in either antiprotons or gamma signals.

Similar considerations apply to cosmic ray positrons, which are
produced by annihilation into leptons (and also by annihilation to EW
gauge bosons). For annihilation into electrons or muons, the constraints
on the DM annihilation cross section from positron observations (for $m_{\rm DM}
\lesssim 100~{\rm GeV}$) are generally significantly stronger than 
those from gamma-ray observations (see e.g.~figure 3 of~\cite{Bergstrom:2013jra}) and CMB perturbations.
For annihilation to taus, the constraints are comparable.

In addition to the limits from indirect detection, models of ADM are
constrained by direct detection experiments and collider searches for
events with missing energy. Specifically, in order for a particle
asymmetry to set the relic density, and thus to obtain ADM, the
symmetric component must annihilate efficiently (with a cross section
larger than that required for symmetric freeze-out) and since these couplings also set
the production and scattering cross sections this can lead to tension
with experimental searches. In particular if the annihilation of the
symmetric component is directly to SM states and can be described via
contact operators (i.e.~from integrating-out some heavy mediator) then
this generally results in fairly strong constraints:
\begin{itemize}
\item Annihilation via contact operators involving SM quarks was studied in \cite{MarchRussell:2012hi} (see also~\cite{Buckley:2011kk}). It was argued that direct detection experiments and LHC monojet limits
typically exclude models of ADM for $m_{\rm DM} \lesssim 100~{\rm GeV}$.
\item LEP searches for mono-photon events \cite{LEPdm} constrain
couplings between DM and electrons. For universal
couplings of DM to charged leptons they exclude the ADM parameter region
for $m_{\rm DM} \lesssim 30~{\rm GeV}$. Preferential annihilation to
$\mu$ or $\tau$ leptons is not constrained.
\item \cite{ColliderEW} considered the limits from collider
observations of DM interactions with electroweak gauge bosons, excluding
ADM models for $m_{\rm DM} \lesssim 40~{\rm GeV}$.
\end{itemize}
These collider limits can be circumvented if the model features a
`light' mediator state \cite{review,MarchRussell:2012hi} (relative to
collider energies i.e.~$\lesssim100$ GeV for LHC searches).
In the remainder of Sect.~\ref{S2}, we study two illustrative examples
in which we can build perturbative models that significantly alleviate these constraints:
\begin{itemize}
\item 
Annihilation of the symmetric component to (on-shell)
metastable vector bosons\footnote{
The simple alternative with a scalar mediator $\phi$ (and $L'$ a
fermion) is not viable, as in this case the annihilation channel $L'
\overline{L'} \rightarrow \phi\phi$ is suppressed by $v^2$. Whilst the
symmetric component can be regenerated through early decays without
conflict with CMB observables, as the DM velocity in the galaxy is
$\sim 10^{-3}$, observably high galactic annihilation rates cannot be
obtained for perturbative couplings in this model.} $L'\overline{L'}
\rightarrow VV$, with $V$ subsequently decaying to SM states (similar to
\cite{Pospelov:2007mp}).
\item Annihilation via the s-channel process  $L'\overline{L'} \rightarrow \phi \rightarrow {\rm SM}$ involving a pseudoscalar mediator $\phi$, with $m_\phi \approx 2 m_{L'}$ such that
the cross section is resonantly enhanced, cf.~\cite{MarchRussell:2012hi}.
\end{itemize}


\subsection{Annihilation via $L' \overline{L'} \rightarrow VV$}
\label{sec:Lpp}

\begin{figure}
\begin{center}
\includegraphics[width=0.6\textwidth]{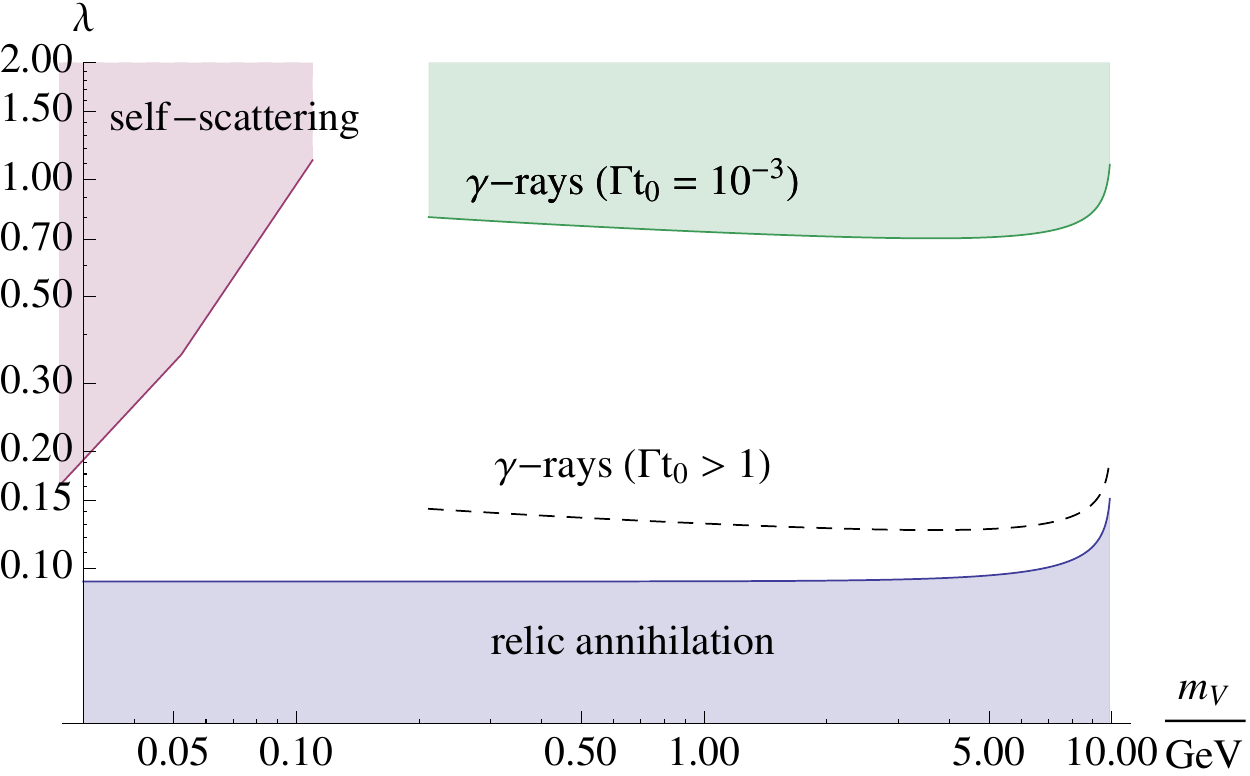}
\caption{Constraints on $\lambda$ for $m_{L'} = 10~{\rm GeV}$, with $V$ decaying to $\mu^+\mu^-$. 
The red curve corresponds to the lower limit on $m_V$ from DM
self-scattering~\cite{Tulin:2013teo}, the blue curve shows the lower
limit on $\lambda$ required for $r_\infty < 0.01$ (i.e.\ for efficient
annihilation of the ADM symmetric component), the dashed black curve
shows the upper bound on $\lambda$ from FERMI gamma-ray bounds assuming that all of
the $B'$ have decayed by the present time ($\Gamma t_0 > 1$), and the green region shows the gamma-ray
bounds for the case where only a fraction $\Gamma t_0 =10^{-3}$ of the $B'$ have decayed.
See Fig.~\ref{fig:regions} for discussion of the gamma-ray bounds.
Decays of $V$ to $e^+ e^-$ would result in broadly similar gamma-ray constraints,
and would allow $m_V$ down to the self-scattering bound (note that, close to the self-scattering bound, the gamma-ray limits would be altered,
since the assumption of a NFW halo is not self-consistent here~\cite{Tulin:2013teo,SelfScatteringHalo}).
}
\label{fig:shell10}
\end{center}
\end{figure}

Let us consider the case that $L'$ is a fermion and there is a the
hidden sector vector boson $V$ which acts as a mediator state. For $m_V
< m_{L'}$, annihilation to a pair of on-shell vectors is possible $L'
\overline{L'} \rightarrow VV$, and if $V$ has a small coupling to the
SM, this channel will dominate.
As an explicit example of this scenario, we consider the following interaction
\beq
\mathcal{L} \supset \lambda \left(\overline{L'} \gamma^\mu L' V_\mu + 
\overline{B'}\gamma^\mu B' V_\mu\right)~,
\eeq
with $V$ subsequently decaying to light SM states. In
Fig.~\ref{fig:shell10} we illustrate how the constraints on $\lambda$
vary with $m_V$, for $m_{L'} \simeq 10~{\rm GeV}$ and assuming that
$V$ decays dominantly to muon pairs (such a coupling can arise due
to kinetic mixing as we discuss in Sect.~\ref{S3}). 
We assume here that the $V$ lifetime is short enough that is does not travel an
astronomically significant distance before decaying --- if it is longer lived,
then the drift of the $V$ from the annihilation point can modify
the observed profile, as discussed in~\cite{Rothstein:2009pm} (though
in that case, we would need to worry about its relic abundance and decays in the
early universe).
The cascade structure of the annihilation means that the resulting
gamma-ray spectrum is softer than that arising from direct annihilations
to muons~\cite{Mardon:2009rc} (as taken into account approximately in Fig.~\ref{fig:shell10}), but since the low-energy part of the spectrum
is important in setting bounds on the annihilation rate at 
DM masses this low~\cite{FERMIdiffuse}, this does not affect the constraints very significantly.
Note that $V$
exchange contributes to DM-DM scattering and, since the self-scattering
cross section increases with decreasing $m_V$, limits on the DM
self-interaction \cite{Tulin:2013teo} (see also \cite{SelfScatteringHalo}) give a lower bound on $m_V$,
as indicated in Fig.~\ref{fig:shell10}. Further constraints on this
example are encapsulated in Fig.~\ref{fig:regions}; for the models
shown the decays are cosmologically slow in order to evade CMB bounds.
Notably, for a wide range of decay rates and mediator masses, models
consistent with experimental constraints exist. 

If $m_V$ is sufficiently small, then the long-range
$V$-exchange between $L'$ and $\overline{L'}$ can give rise to Sommerfeld-type
low-velocity enhancement of the annihilation rate. However, the lower
limit on $m_V$ from self-scattering implies that this is not significant here.
Also, as discussed in Sect.~\ref{sec:splitting}, DM self-scattering cross sections
near to the observational limit (as sometimes invoked to solve problems with small-scale structure
formation~\cite{Spergel:1999mh}) have consequences for the shape of the DM halo, 
though we do not consider the details here.


\subsection{Annihilation via $L' \overline{L'} \rightarrow \phi \rightarrow {\rm SM}$}
\label{sec:schanlim}

We now turn to the second model, in which $L' \overline{L'}$ annihilation
proceeds through the s-channel exchange of a mediator $\phi$, where
$m_\phi$ is close to resonance.
For the case where $m_\phi > m_{L'}$, the $L'\overline{L'}$
annihilations cannot produce on-shell $\phi$ pairs and rather
proceed directly to SM states via a $\phi$-mediator. Since we expect
$L'$ to be roughly GeV or greater, the DM-proton scattering cross
section is well described by an effective operator and if $\phi$
is a scalar or a vector, coupling to quarks, then large regions
of parameter space are excluded by direct detection experiments,
see~\cite{MarchRussell:2012hi}.\footnote{More complicated
portal interactions can alleviate some direct detection constraints,
e.g.~\cite{Belanger:2013tla,DelNobile:2013gba,Fox:2013pia,Gresham:2013mua}}
Whilst the effective theory is valid
for DM-proton scattering, kinematic effects can be important to the
annihilation process and can potentially lead to regions of parameter
space which are not excluded by direct searches, as we discuss below.

Given $\phi$-DM and $\phi$-SM couplings, the 
$L'\overline{L'} \rightarrow {\rm SM}$ annihilation cross section is
dramatically increased if $m_\phi \approx 2 m_{L'}$, such
 that the s-channel annihilation is close to resonance. 
 This does not make the direct search constraints
tighter, so gives us more parameter space for the model. In particular,
if $m_\phi$ is slightly above $2 m_{L'}$, then at higher temperatures
the thermal averaging samples more of the resonance peak, increasing
the annihilation cross section. Effectively an
enhancement of $\langle \sigma v\rangle$ at high temperatures occurs, as
required for the early-decay case to be viable.

\begin{figure}
\begin{center}
\includegraphics[width=0.6\textwidth]{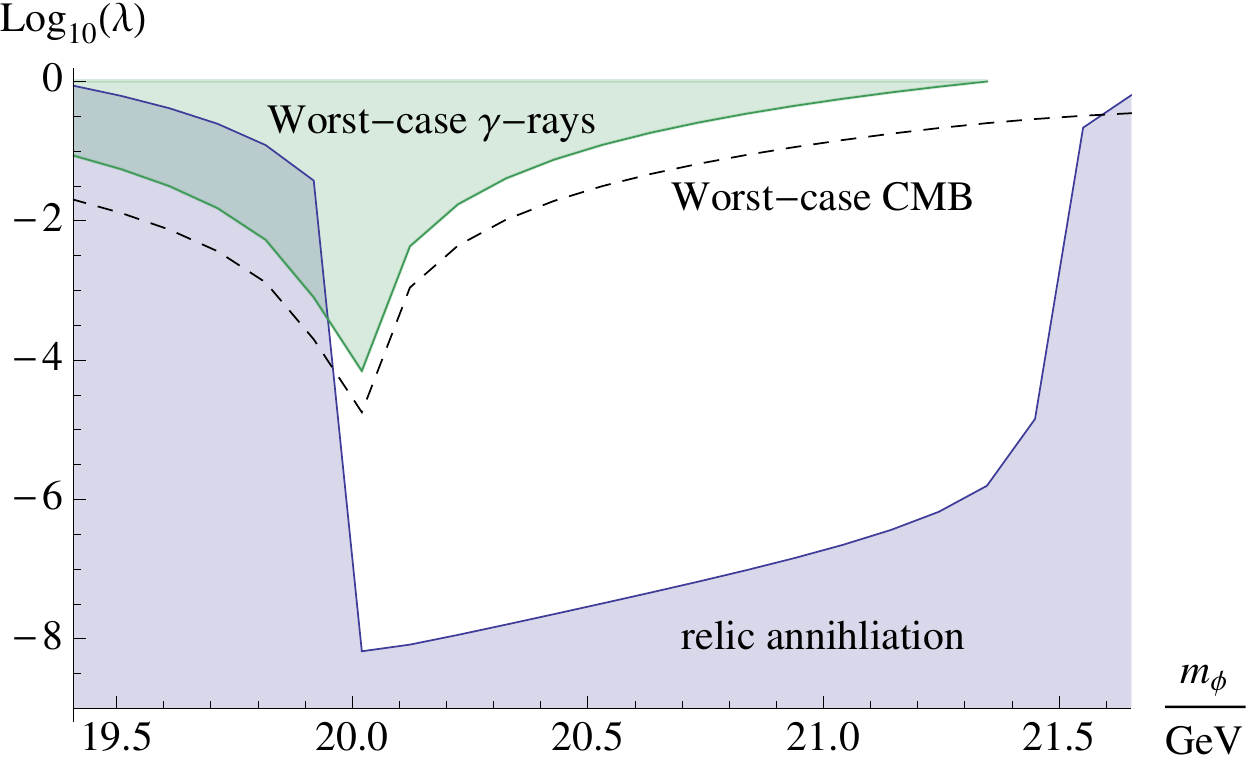}
\caption{Constraints on $\lambda$ for $m_{L'} = 10~{\rm GeV}$,
for coupling to quarks as described in section~\ref{sec:schanlim}. The
blue curve shows the lower limit on $\lambda$ required for $r_\infty <
0.05$ (i.e.\ for efficient annihilation of the ADM symmetric component),
and the green region is excluded by
gamma-ray bounds (see Fig.~\ref{fig:regions} for details), assuming that all of the $B'$ have decayed by the present time.
The black dotted curve shows the exclusion bound from CMB perturbations,
assuming that all of the $B'$ have decayed by recombination time.
The relic abundance and annihilation cross-section calculations were carried out using the
{\tt micrOMEGAs} package~\cite{Belanger:2001fz}.
}
\label{fig:res10}
\end{center}
\end{figure}

To illustrate the allowed parameter space around the resonance region, we
consider an explicit model with a pseudoscalar $\phi$ coupling to quarks via
\begin{equation}
\mathcal{L} \supset i \lambda \phi \overline{L'} \gamma^5 L' + \sum_q i \lambda' y_q \frac{m_\phi}{m_h} \phi	\overline{q} \gamma^5 q~.
\end{equation}
Fig.~\ref{fig:res10} shows the constraints in the $(m_\phi, \lambda)$
plane for $\lambda' = 0.14$ (the behaviour is mostly dependent on $\lambda \lambda'$, so we only plot variation  with $\lambda$). As discussed above, there is a strong distinction between
$m_\phi$ just above vs below $2 m_{L'}$. Above, the enhancement of the annihilation cross section
with temperature means that there is a large allowed region between the couplings
necessary for relic annihilation and those ruled out by astrophysical constraints. For $m_\phi < 2 m_{L'}$, on the other hand, higher collision
energies result in $\phi$ being more off-shell. Consequently, in this region, the relic
annihilation constraint rules out everything until $\lambda$ is very large.

The figure shows the strongest bounds, when the decays are early enough
to maximise the limits from both CMB and galactic timescales. Since
this still leads to viable parameter space, early decays are permitted
(unlike the previous models, which as seen in Fig. \ref{fig:regions}
are very constrained if there are early decays). Note that, as we increase
the decay time past recombination time, the CMB limits become less constraining
relative to the gamma-ray bounds (cf.\ Fig.~\ref{fig:regions}), so
(depending on assumptions
about antiproton propagation, cf. Fig~\ref{fig:pbar}) there 
is a large parameter space in which gamma-ray signals may realistically be seen before
other signals.

\subsection{Tentative signals and morphology}
\label{3.6}

Recently, there have been suggestions \cite{Hooper:2013rwa} (see also \cite{GC}) of a gamma-ray excess from
the galactic centre and its vicinity, compatible with $\sim 10~{\rm
GeV}$ DM (NFW-profile) annihilating to leptons (with $\langle
\sigma v\rangle \simeq 2 \times 10^{-27} {\rm cm}^3 {\rm s}^{-1}$)
or with $\sim 50~{\rm GeV}$ DM annihilating to quarks (with $\langle
\sigma v\rangle \simeq 8 \times 10^{-27} {\rm cm}^3 {\rm s}^{-1}$). Both models discussed above can
reproduce a signal of this kind, but only with a sufficiently small $B'$-$L'$ mass
splitting such that the $\overline{L'}$ profile remains sharply peaked towards the galactic
centre. It was claimed in \cite{Hooper:2013rwa} that, for symmetric annihilating
DM, the signal is fit well by a generalised NFW type profile with shape
parameter $\gamma \simeq 1.2$, and with some systematic uncertainty possibly
allowing $\gamma$ to vary around this value to a maximum of $\sim 2$ (see also~\cite{Gordon:2013vta}).

\begin{figure}
\begin{center}
\includegraphics[width=0.49\textwidth]{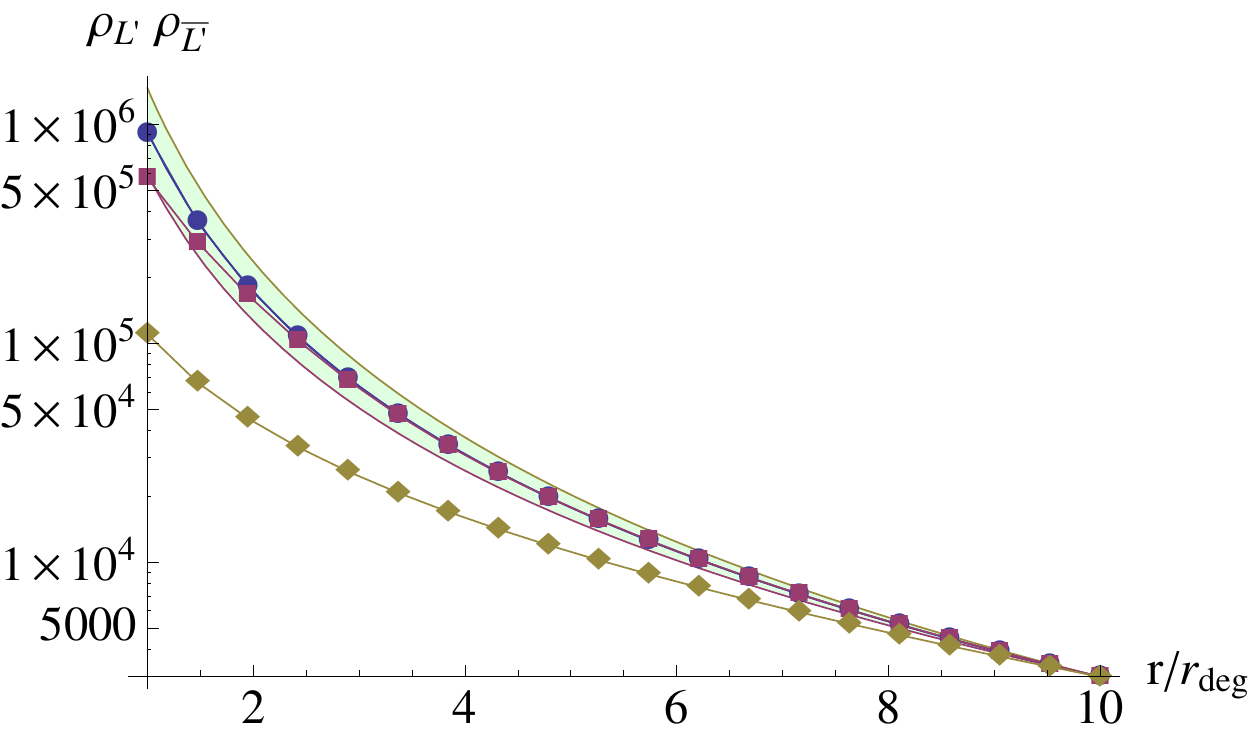}
\includegraphics[width=0.49\textwidth]{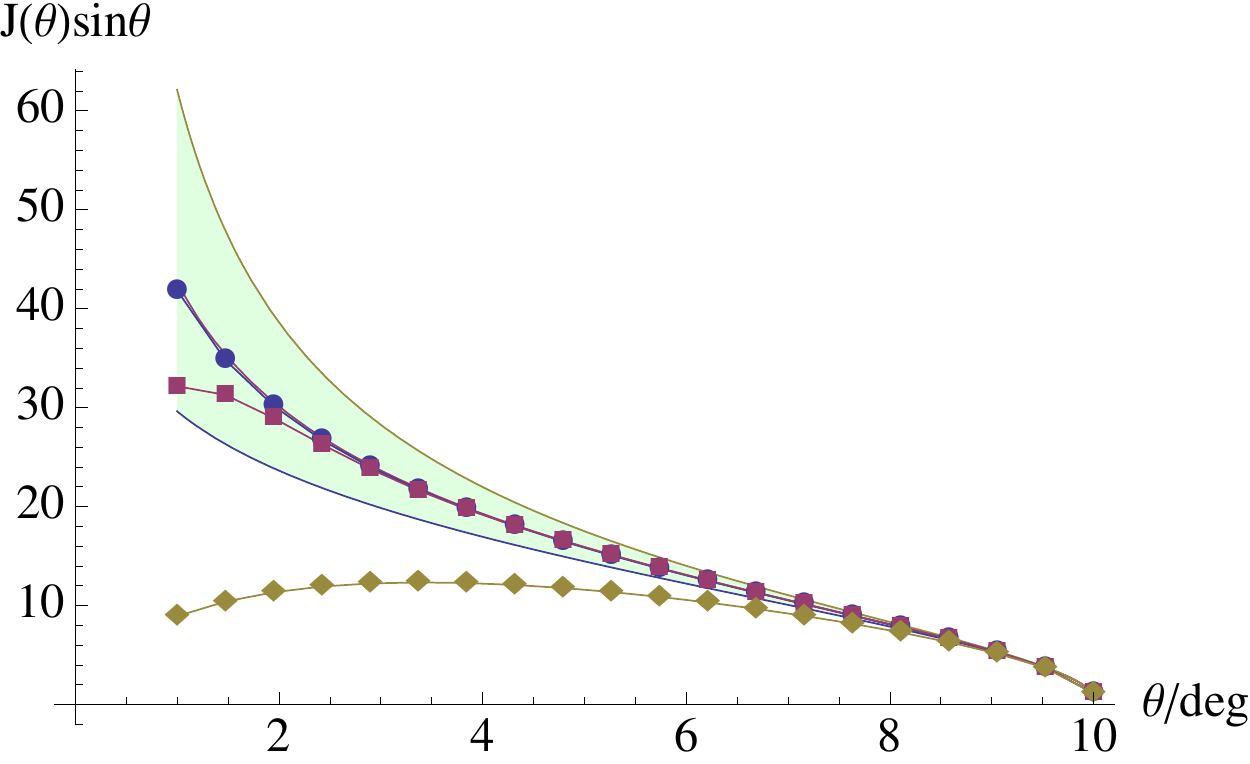}
\caption{
\textbf{\em Left:} The $\rho_{L'} \rho_{\overline{L'}}$ profile for different $L'$ and
$\overline{L'}$ profiles (in arbitrary units, normalised to a common
value at $r = 10 r_{\rm deg}$, where a distance of $r_{\rm deg}$ above the galactic centre corresponds to $b = 1^\circ$). 
The blue dots indicate the $\gamma=1.2$
$L'$ profile, and a $\overline{L'}$ profile obtained by starting with a NFW
$\gamma = 1.2$ profile and applying a velocity kick of $1~{\rm km}{\rm
s}^{-1}$. The red and yellow dots correspond to the same situation, but
with velocity kicks of $76~{\rm km}{\rm s}^{-1}$ and $163~{\rm km}{\rm
s}^{-1}$ respectively. The shaded region corresponds to generalised NFW profiles (for $L'$ and $\overline{L'}$) with $1.1 < \gamma < 1.3$ required to best match the apparent gamma-ray 
excesses near the galactic centre  \cite{Hooper:2013rwa}.
\textbf{\em Right:} For a given model the signal size is determined by the integral $J(\theta)=\int dl \rho_L\rho_{\overline{L}}$ along the line-of-sight as a function of angle from galactic centre. Colours and shading as in left panel.}
\label{fig:nfwHooper}
\end{center}
\end{figure}

Fig.~\ref{fig:nfwHooper} shows the profile shapes resulting from
different $B' \rightarrow \overline{L'} + \cdots$ injection velocities, starting from
a NFW-type ($\gamma=1.2$) $B'$ profile. These are calculated as described in Appendix~\ref{ApB},
utilising the properties of the steady-state DM phase space distribution function. As it illustrates, we need $v_i
\lesssim 120~{\rm km}{\rm s}^{-1}$ to match a symmetric annihilation
signal with $1.1 < \gamma < 1.3$, corresponding to a fractional
$B'$-$L'$ mass splitting of $\lesssim 4 \times 10^{-4}$ (subject to the caveats
mentioned in Sect.~\ref{sec:splitting} and Appendix~\ref{ApA}). If the mass
splitting is not much smaller than this limit, then the DM gamma-ray signals from structures
with lower velocity dispersions than the galaxy will be modified
compared to what we expect from standard annihilating symmetric DM.
At present, the nature of these structures (e.g. DM clumps in the
galactic halo) is poorly constrained, so it is hard to pick out any
definite differences, but since the escape velocity for most dwarf galaxies
is below $\sim 50~{\rm km}{\rm s}^{-1}$~\cite{Dehnen:2005fs}, and the velocity dispersion is $\sim 10~{\rm km}{\rm s}^{-1}$~\cite{Walker:2007ju}, if $v_i$ is high enough then we will almost completely
suppress the annihilation signal from dwarf galaxies due to most of the
$\overline{L'}$ population escaping. However, with present and
near-future experiments, gamma-ray observations of dwarf galaxies
place significantly weaker constraints on DM annihilation cross sections
than galactic centre observations~\cite{dSphGamma}, without the assumption of
large boost factors due to DM clustering or velocity suppression. As a
consequence, for annihilation cross sections at the sub-thermal level
required to replicate the signal under discussion, we would not expect
to see the corresponding signals from dwarf galaxies in the near future.

In addition to the modified $\overline{L'}$ distribution in space, the main
possibly-observable difference from a standard annihilating symmetric
DM scenario is the change in $\overline{L'}$ population over time. However, due to
the small (symmetric) annihilation cross section required to match the
signal, the effect on the CMB from symmetric annihilations would
be below cosmic variance~\cite{CMB}, so it would not be possible to detect
the difference between this and a smaller effect in the $\overline{L'}$ case.
This is not a completely general statement --- there do exist viable symmetric DM
models which have observable annihilation effects on the CMB, and
in these cases the late-decay ADM model would make different predictions.
Also note that possible collider signals of an ADM model may differ
from those of a symmetric DM model (though in neither case are we guaranteed
to have such signals, e.g.\ the models of Sect.~\ref{sec:Lpp} in our case, and
the symmetric model of~\cite{Boehm:2014hva}).


\section{Model building}
\label{S3}

It was argued in the previous section that a small mass splitting is a
natural way to obtain observable annihilation signals. Hence, next
we highlight a motivated setting in which such a scenario may occur.
Specifically, we outline a model based on broken
flavour symmetries in the hidden sector. Further, drawing on analogies with nuclear
(proton-neutron) mass splitting, we sketch an alternative realisation of such a set-up in models of composite DM.


\subsection{An implementation with fundamental matter}
\label{S3.1}

We first present a simple implementation involving (fundamental) matter
that obtains Dirac masses, with a small mass splitting generated from
radiative breaking of a flavour symmetry. The hidden sector has a $Z_2$
symmetry, an exact $B-L-(B'-L')$ symmetry, and accidental approximate
$B'$ and $L'$ symmetries.\footnote{The couplings studied here actually
respect $B'-L'$, however once a mechanism to cogenerate or share
asymmetries with the visible and hidden sectors is included, the true
symmetry is $B-L-(B'-L')$.} The matter content of our (example) model
is displayed in Tab.~\ref{tab:t1}; it features a heavy complex scalar $\phi$
and fermion matter, which is written as two component left-handed Weyl
spinors. The typical mass hierarchy of the states is displayed to the
right of Tab.~\ref{tab:t1}.

\begin{table}
\begin{multicols}{2}

\begin{tabular}{| c | c ||  c || c | c | c | c |}
    \hline
   & spin   & $Z_2$  & $B'$& $L'$ &$B'-L'$ \\ \hline
       $\phi$ & 0  & $1$  &  $-$ & $-$ & $0$ \\    \hline
       $B'{}_{L\alpha}$ & $\frac{1}{2}$  & $1$ &  $1/2$ & $0$ & $1/2$ \\ \hline
       $B'{}_{R\alpha}^\dagger$ & $\frac{1}{2}$  & $1$  &  $-1/2$ & $0$ & $-1/2$ \\ \hline
       $L'{}_{L\alpha}$  &$\frac{1}{2}$  & $0$ & $0$ & $1/2$& $-1/2$ \\ \hline
       $L'{}_{R\alpha}^\dagger$ & $\frac{1}{2}$  &  $0$ & $0$ & $-1/2$& $1/2$ \\ \hline
       $N_{1\alpha}$ & $\frac{1}{2}$ &$1$  &  $0$ & $0$ & 0 \\ \hline
       $N_{2\alpha}$ & $\frac{1}{2}$ & $0$ & $0$ &  $0$ & $0$ \\ \hline
    \end{tabular}

\columnbreak
\hspace{15mm}
\includegraphics[width=35mm]{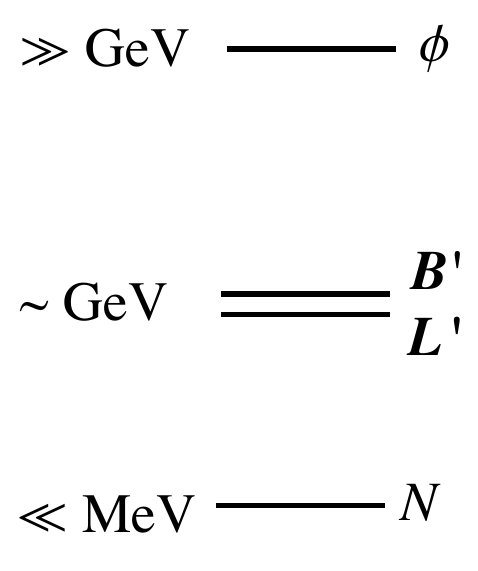}

\end{multicols}
\caption{\textbf{\em Left:} hidden sector fundamental matter content. \textbf{\em
Right:} hidden sector mass hierarchy.}
\label{tab:t1}
\end{table}

\begin{figure}[t!]
\begin{center}
\includegraphics[width=0.45\textwidth]{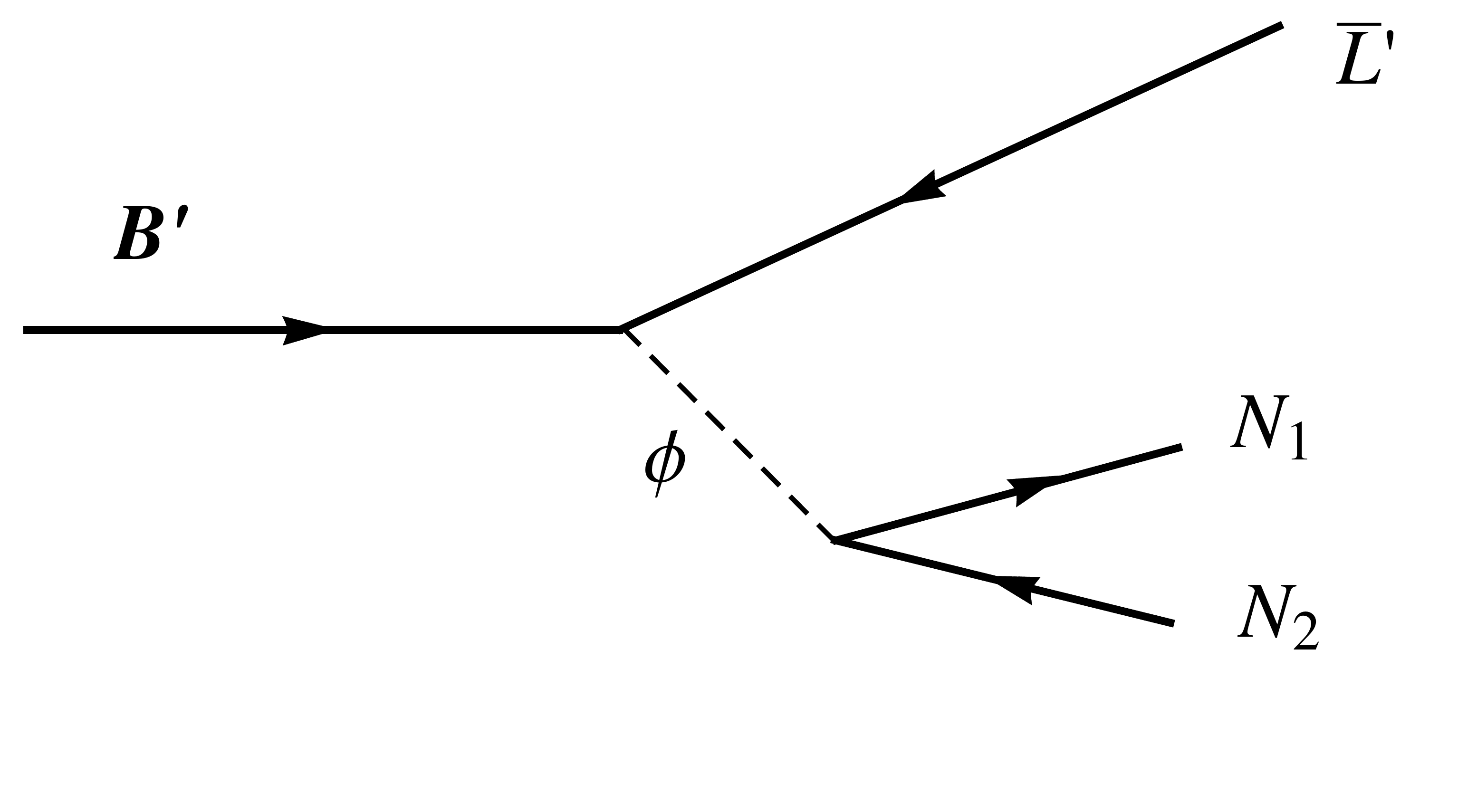}
\caption{$\overline{L'}$ are generated via $B'$ decays and then annihilate with the asymmetric population of $L'$.}
\label{fig:fundecay}
\end{center}
\end{figure}

The Lagrangian involving the light fields can be expressed as
\begin{equation}
\mathcal{L} \supset -m_{B'} B'_{L \alpha} B'{}^{\alpha~\dagger}_R -  m_{L'} L'_{L \alpha} L'{}^{\alpha~\dagger}_R + {\rm h.c.}
\end{equation}
Note that Majorana mass terms for $L'$, $B'$ are forbidden by the exact $B-L-(B'-L')$ symmetry and mixing between these states is forbidden by the $Z_2$ symmetry. It can be seen that $B'$ and $L'$ arise as approximate symmetries since there are no renormalisable terms in the low energy effective theory  that violate them. However, the heavy scalar $\phi$ has couplings of the form
\begin{equation}
\mathcal{L} \supset  y_1 \phi \left( B^{\alpha}_L L_{L\alpha} + B_{R \dot{\alpha}} L_R^{\dot{\alpha}}  \right) + y_2 \phi N_1^{\alpha} N_{2 \alpha} + \rm{h.c.}
\end{equation}
As a result, this can mediate decays of $B'$ to $\overline{L'}$ as shown
in Fig.~\ref{fig:fundecay}. We shall assume that the scalar $\phi$ gains
a mass well above $B'$ and $L'$ and hence can be integrated out, leading
to a four-fermion operator. 
The lifetime of $B'$
is given in eq.~(\ref{lifetime}), setting $M_{\not B'}=m_\phi$ and
hence, for a given model, one can compare with the limits indicated in
Fig.~\ref{fig:regions}.

The fields $N_1$ and $N_2$ are assumed to gain Majorana masses $M_1$ and $M_2$, which are parametrically smaller than the $B'$ and $L'$ (Dirac) mass scale\footnote{
As mentioned in Appendix~\ref{app:rates}, if $N_1$ and $N_2$ do not have other interactions,
then their masses must generally be ($< \mathcal{O}({\rm eV})$) to avoid having too large a relic density.}. 
\begin{equation}
\mathcal{L} \supset -M_{1} N_{1 \alpha} N^{\alpha}_1 -M_{2} N_{2 \alpha} N^{\alpha}_2 + {\rm h.c.}
\end{equation}

Further, we assume that $m_{B'} =m_{L'}$ at leading order, exhibiting an approximate `flavour' symmetry which is only radiatively broken. This breaking could be due to the differing $Z_2$ parities of these states, resulting in the $B'$ and $L'$ having different couplings to some additional (possibly heavy) matter content in the theory. Alternatively, we could introduce an additional gauge group to the low energy theory, under which the $B'$ and $L'$ have differing charges. Such a soft breaking of the $Z_2$ flavour symmetry will lead to a small mass splitting between $B'$ and $L'$, similar to that employed in models of inelastic DM \cite{TuckerSmith:2001hy} and eXcited DM \cite{Finkbeiner:2007kk}. 

To provide a portal to the SM, we can introduce an additional gauge
boson, $Z'$, under which the states $L'$ (and $B'$) are charged.
Provided this has the appropriate mass (through an additional scalar,
that we do not specify, gaining a vacuum expectation value) annihilation
of $L'$ and $\overline{L'}$ proceeds to two on-shell $Z'$. If the
states $N_1$ and $N_2$ are uncharged under this symmetry, and there
are no other lighter hidden sector states to which the $Z'$ can decay,
then it will be approximately stable. Decay of the $Z'$ to the SM can
then occur through, for example, a small amount of kinetic mixing with
the SM hypercharge U(1). This is particularly well motivated from the
perspective of a string theory UV completion. In a IIB model, the
hidden sector can arise as a theory on branes at a singularity of the
compactification that is geometrically separated from the SM branes. If
the distance between the two sectors is large, there are typically no
couplings in the low energy theories, except for kinetic mixing which
is unsuppressed by such a separation \cite{Abel:2008ai}. Alternatively,
it is straightforward to introduce an additional pseudoscalar in the
theory that couples to the states $L'$. This can then act as a portal
if it also couples to the SM quarks, leading to a model of the form of
Sect.~\ref{sec:schanlim}.

While in the model presented here the decay of $B'$ to $\overline{L'} + \cdots$ is mediated through a heavy scalar, there are also well motivated scenarios where this occurs through a heavy vector. In particular, if $B'$ and $\bar{L'}$ appear in the same multiplet in a (spontaneously broken) GUT theory, this decay may be mediated through a gauge boson charged under the $Z_2$ symmetry, analogous to the X, Y bosons that appear in Standard Model GUTs. In this case, the $Z_2$ symmetry must also arise as a discrete remnant of part of the hidden sector GUT gauge group, so that the states $B'$ and $\bar{L'}$ can have differing charges under this.


\subsection{Comments on composite-type models}
\label{S3.2}

Alternatively, a small mass splitting between states can be accomplished
in models of composite DM, in analogy to the proton-neutron mass
splitting.
The phenomenology of such composite models is, while potentially
interesting, generally rather complicated, and it is not obvious that
all of the necessary constraints can be satisfied in a simple model.
Nevertheless, we sketch one such model below, without investigating
phenomenological details.

Consider a simple hidden sector with a confining gauge group, say
SU(2), and left-handed chiral matter content consisting of $Q'_{\alpha}$
and $E'_{\alpha}$.
Suppose the hidden sector SU(2) runs into strong coupling at about
$5\,\rm{GeV}$, then the theory will confine. As a result calculations
of the dynamics in this region are unreliable, hence, we shall simply
assume that the lowest mass states which arise as $\boldsymbol{2}\times
\boldsymbol{2}$ are $E' E'$ bound states, and identify these with
$L{^\prime}$, and the next lightest is the $Q' Q'$ composite which we
call $B^{\prime}$, as given in Tab.~\ref{T2}.
There are also mixed meson states
e.g.~$\overline{Q'}Q',~\overline{E'}E',~\frac{1}{\sqrt{2}}\left(Q' E'
\pm E'Q'\right)$, which unfortunately complicate the phenomenology.

\begin{table}[h!]
\begin{center}
    \begin{tabular}{| c | c | c | c | c | c | c |}
    \hline
     &constituents & $B'$& $L'$ &$B'-L'$ \\ \hline
    $B^{\prime}$ & $Q'Q'$ & $1$ & $0$ & $1$ \\ \hline
    $L^{\prime}$ & $E'E'$   &$0$ & $1$& $-1$ \\ \hline
    $\pi^\prime$ & $\overline{Q}Q,~\overline{E}E$ & $0$ & $0$ & $0$ \\ \hline
    \end{tabular}
    \caption{ \label{T2} Lightest bound state singlets of the hidden
sector gauge group (plus their anti-partners.)}
\end{center}
\end{table}

Again, the theory is assumed to have an exact  $B - L - (B' -L')$ symmetry, which leads to approximate $B'$ and $L'$ symmetries in the low energy effective field theory. 
The mass splitting between $B'$ and $L'$ can arise due to radiative
corrections, as in the previous `flavour' example. For instance, suppose
the states $Q'$ and $E'$ are charged under an additional hidden sector
gauge group; then the typical size of the $B'$-$L'$ mass splittings
due to these extra gauge interactions (with coupling constant $g$) is
parametrically \cite{ArkaniHamed:2008qn,Baumgart:2009tn}
\begin{equation}
\Delta m \sim \frac{g^2}{16 \pi^2} \Lambda~,
\end{equation}
where $\Lambda$ is the confining scale of the hidden SU(2).

Further complications arise due to the composite nature of the DM,
in particular the annihilation and scattering cross sections can be
modified, possibly involving intricate form factors, and it is likely that
`dark pion' $\pi'$ exchange will lead to DM self interactions fairly close to
current limits. Notably, this potentially also allows for enhancements
in the annihilation rates which determine the relic density or indirect
detection signals. For some related discussions of composite DM see
e.g.~\cite{comp}. Thus, whilst composite models are an interesting setting 
for realising a small mass splitting, given these complications we shall 
not pursue this scenario in further detail here.


\section{Conclusion}
\label{S4}

ADM is a well motivated framework in which to examine proposals
of light DM, as if the baryon and DM asymmetries are comparable,
then the fact that $\Omega_{\rm{\rm DM}}\sim5\Omega_{B}$ implies that the DM mass is similar to the proton. Annihilation-like 
indirect detection signatures are typically not expected if the DM relic
density is determined by a particle asymmetry, but can arise in models with more complex hidden sectors. We have outlined a 
new class of ADM models in which annihilation signals are
 generated, and presented a number of concrete examples. 
In particular we discussed a scenario involving two states $B',~L'$ with comparable relic densities, stabilised by two different approximately conserved quantum numbers. Subsequently, processes which violate these approximate symmetries lead to decays of the heavier state $B'$, regenerating the symmetric component of the lighter species $L'$.  This allows for the prospect of observable indirect detection signals with annihilation-like profiles via $L'$-pair annihilation.

One of the principal model-building challenges for producing observable
signals in this manner is that, unless the mass splitting between $B'$
and $L'$ is small, most of the $\overline{L'}$ generated via $B'$
decays are immediately ejected from the galaxy and do not give rise
to annihilation-type signals. The desire for observable annihilation
signals constrains the parameter space of the model as discussed in
Sect.~\ref{sec:splitting} and illustrated in Fig.~\ref{fig:nfwAlpha}.
Moreover, if the mass splitting is moderate then the symmetric component
of $L'$ may escape galactic structures with low escape velocities. Thus
a distinctive signature of this class of models is that annihilation
signals could be observed in our galaxy and conspicuous by their absence
in dwarf galaxies. Further, we outlined two scenarios in which such
small mass splittings can arise, in the context of broken flavour
symmetries, and composite models analogous to the proton-neutron mass
splitting.

There is a range of further experimental constraints on these models both from direct and indirect probes, as encapsulated in Figs.~\ref{fig:regions}, \ref{fig:decSM}, \ref{fig:shell10}. In order to evade the strong direct detection bounds we studied two particular scenarios: in the first case annihilations occur on resonance and in the alternative the DM annihilates to pairs of meta-stable hidden sector states. We highlighted the prospect for indirect detection signals in the near future, see for instance Figs.~\ref{fig:regions} \& \ref{fig:pbar}, and commented on tentative signals of DM annihilations near the galactic centre \cite{Hooper:2013rwa,GC} in Sect.~\ref{3.6}.

To conclude, we find that whilst models of low-mass ADM with a symmetric component regenerated by decays can produce observable annihilation signals, and satisfy the various experimental bounds, these models are typically required to possess some
specific properties and thus exhibit some predictive features. 
Several experiments have hinted at the possibility of DM in the 1-50 GeV
range, both direct detection \cite{DD} and
 indirect signals of DM annihilation \cite{Hooper:2013rwa,GC}.
The confirmation of annihilation signals consistent with light DM
(possibly with an annihilation cross section different from the thermal
freeze-out value) could be an indication of the class of annihilating
ADM models proposed here.


\section*{Acknowledgements}

We are grateful to Ulrich Haisch, Dan Hooper, Felix Kahlhoefer, John
March-Russell, Chris McCabe, Michele Papucci, and especially Stephen West for useful discussions. EH and RL are
funded by STFC Studentships. In addition, RL is grateful for support
from the Buckee Scholarship at Merton College, Oxford.


\appendix


\section{Rates of $B'$ and $L'$ violating processes}
\label{app:rates}

In this work, we have assumed that all $B'$ and $L'$ violating processes
apart from the (slow) $B' \rightarrow \overline{L'} + \cdots$ decay can be neglected.
Here, we justify this assumption and estimate the rates of various processes.

Suppose that $B'$ and $L'$ are good symmetries at low energies,
and are only violated at some high scale $M_{\not
B'}$. The decay of $B'$ is then described by a contact operator
$\frac{1}{\Lambda^k}B'L'\mathcal{O}_{k+1}$, where we assume that $B'$
and $L' $ are fermions, and so $\mathcal{O}_{k+1}$ is a dimension $k+1$
operator. The small mass splitting $\Delta m$ between $B'$ and $L'$ will
result in a phase-space suppression of the decay width, and there will
be
additional $\Delta m$ factors from any fermionic wavefunctions from
$\mathcal{O}_{k+1}$. Dimensionally, a dim-($4+k$) operator will give a
total width going parametrically as $(\Delta m)^{2k + 1}$
(or $(\Delta m)^{2k+3}$ if the $B'L'$ current gives another $(\Delta m)^2$).
For example, a dimension-6 operator will, in the former case, give a total width of
\beq
\Gamma \sim \frac{(\Delta m)^5}{\Lambda^4}~.
\eeq
In contrast, symmetry-violating annihilations (for instance $B' + L' \rightarrow
\cdots$) at high temperatures will not be suppressed, and for a dim-($4 + k$)
operator will have $\sigma \sim \frac{T^{2k - 2}}{\Lambda^{2 k}}$.
Hence, in the early universe, the rate of such interactions will be 
$n_{L'} \langle \sigma v\rangle \sim \frac{T^{2k + 1}}{\Lambda^{2 k}}$.
So, for $T$ much larger than $\Delta m$, there is the possibility that 
these will be fast enough to affect the asymmetries in $B'$ and $L'$ (if these
are established above that temperature).

Specifically, since $B' - L'$ is a good symmetry (ignoring processes that
only respect the full $B - L - (B'-L')$), we will have $\eta_{B'} - \eta_{L'} = {\rm constant}$.
By symmetry, the Boltzmann equation for the other linear combination $\eta_{B'} + \eta_{L'}$
is
\begin{equation}
\frac{d(\eta_{B'} + \eta_{L'})}{dx} = - \lambda(x) (\eta_{B'} + \eta_{L'})~,
\end{equation}
to leading order in the small asymmetries, where $x \propto 1/T$. So,
if $\lambda$ is large enough, the asymmetries will be driven towards
$\eta_{B'} = - \eta_{L'}$, corresponding to a population of $B'$ and
$\overline{L'}$ (or the reverse, depending on the sign of $\eta_{B'}
- \eta_{L'}$). This would not permit the scenarios we discussed in
the main text, where the asymmetries have the same sign. It may be
possible to build models in which $B'$ and $\overline{L'}$ populations
give rise to the decay-followed-by-annihilation signals considered in
the phenomenological sections of this paper (for example, through $B'
\rightarrow L' + \nu\nu + \cdots$ type decays which conserve $B - L -
(B' - L')$), but we do not go into any model-building details here.

To check more quantitatively whether this asymmetry-reversal will
be important, we take as an example the model
of Sect.~\ref{S3.1}, which has (schematically) the symmetry-violating operator
$\frac{1}{\Lambda^2} (B'L')(N_1^\dagger N_2^\dagger)$.
This leads to a three-body decay, with differential width
\begin{equation}
d\Gamma = \frac{1}{2 m_{B'}} |\mathcal{A}|^2 dq^2 \frac{d\Omega_{\rm CM}}{16 \pi^2} \frac{|\vec{p}_{\rm CM}|}{m_{B'}}\frac{1}{2 \pi}\frac{d\Omega_N}{16\pi^2}\frac{|\vec{p}_N|}{\sqrt{q^2}}~,
\end{equation}
where $q$ is the total 4-momentum of $N_1$ and $N_2$, and the subscript-$N$ quantities are
in the frame where $q$ is purely timelike. Neglecting the masses of $N_1$ and $N_2$, and
assuming that $\Delta m$ is small, we have ${\vec{p}_{\rm CM}}^{~2} \approx (\Delta m)^2 - q^2$.
Also, $|\mathcal{A}|^2 \approx \frac{4}{\Lambda^4} m_{B'}^2 q^2$ (summing over final spins), so overall,
\begin{align}
\Gamma &\approx \frac{1}{4 \pi m_{B'}} \int_0^{(\Delta m)^2}dq^2\, \frac{1}{4\pi}
\frac{\sqrt{(\Delta m)^2 - q^2}}{m_{B'}} \frac{1}{8 \pi} \frac{4}{\Lambda^4}m_{B'}^2 q^2
\\
&= \frac{1}{(4 \pi)^3}\frac{8}{15}\frac{(\Delta m)^5}{\Lambda^4}~.
\end{align}

In the early universe, symmetry-violating processes of the form
$B' + L' \rightarrow N_1 + N_2$, $B' + N_1 \rightarrow \overline{L'} + N_1$, etc.~will be active.
The associated annihilation cross sections are
\begin{equation}
\sigma \approx C \frac{1}{8 E^2} \frac{1}{8 \pi} \frac{1}{\Lambda^4} E^4~,
\end{equation}
where $E \gg m_{B'}$ is the energy of each particle in the CoM frame, and $C$ is a numerical
constant depending on which legs are ingoing and outgoing. 
Thus, in a thermal bath at temperature $T \gg m_{B'}$, we have the thermally averaged cross
section~\cite{Gondolo:1990dk}
\begin{equation}
\langle \sigma v \rangle \approx \frac{3}{32 \pi} C \frac{T^2}{\Lambda^4} \equiv C \sigma_1 \frac{T^2}{T_1^2}~,
\end{equation}
where $T_1$ is some (high) temperature.
The Boltzmann equation is then (to leading order) of the form
\begin{equation}
\frac{1}{a^3}\frac{d}{dt}\Big[a^3\big((n_{B'} - n_{\overline{B'}}) + (n_{L'} - n_{\overline{L'}})\big)\Big]
= - C' n_{N_1}(T) \sigma_1 \frac{T^2}{T_1^2} \Big[(n_{B'} - n_{\overline{B'}}) + (n_{L'} - n_{\overline{L'}})\Big]~,
\end{equation}
where $C'$ is a numerical constant obtained from summing over all of the
leg orderings, with the appropriate weights (its value is $\mathcal{O}(10)$).
Converting to conserved variables $Y_i = n_i/s$, and letting $x = T_1/T$, this becomes
\begin{equation}
\frac{d(\eta_{B'} + \eta_{L'})}{dx} = -\frac{\lambda}{x^4} \sqrt{g_*(T)} Y_{N_1}(\eta_{B'} + \eta_{L'})~,
\end{equation}
with
\beq
\lambda \simeq  1.32 \times C' T_1 M_{\rm Pl} \sigma_1~,
\eeq
where we have used that during the radiation-dominated era, the Hubble
rate is given by $H = \frac{T^2}{M_{\rm Pl}^*}$, where $M_{\rm Pl}^* =
M_{\rm Pl} \sqrt{\frac{90}{\pi^2 g_\star(T)}}$. So, in terms of the $B'$
decay rate $\Gamma$, we find that
\begin{equation}
\lambda = 3 \pi^3 \sqrt{\frac{5}{2}} C' \frac{T_1^3 M_{\rm pl} \Gamma}{(\Delta m)^5}~.
\end{equation}
So, if $T^3 \gtrsim 10^{-3} \frac{(\Delta m)^5}{\Gamma M_{\rm pl}}$,
then $\eta_{B'} + \eta_{L'}$ will be suppressed by multiple $e$-folds.
Taking some representative values, for $\Gamma \sim 1/t_0$ and
$\Delta m \sim 10\,{\rm MeV}$, we find that this corresponds to $T
\gtrsim 3\,{\rm TeV}$. For the models proposed here to be successfully
realised, it is required that the $B'$ and $L'$ asymmetries are
established at some point after the universe has cooled below this
temperature.

Further, our assumption that the asymmetries are set before $B'$,
$L'$ freeze-out from the thermal bath, i.e.\ roughly before they
become non-relativistic, implies a lower limit on $\Delta m$, or
equivalently, an upper limit on $\Gamma$. For example, for $\Gamma =
1/t_{\rm CMB}$ it is required that $\Delta m \gtrsim 7\,{\rm MeV}$,
otherwise symmetry-violating interactions do not decouple until $T
\lesssim 50\,{\rm GeV}$. This limit is plotted in Fig.~\ref{fig:decSM},
demonstrating that there is viable parameter for all cosmologically
slow decay rates.

There is also the possibility of symmetry-violating interactions at
late times, but the number densities then are small enough to make
these completely negligible. The most frequent will be those involving
a $N$ particle, since (in the simplest case) these decouple from the thermal bath when the
symmetry-violating interactions discussed above decouple, forming a dark
radiation component (the large difference between $g_*$ at the time of
decoupling and later means that this does not conflict with constraints
on $N_{\rm eff}$ from BBN and the CMB). The $N$ will have a number density
of $\sim n_\gamma/8$, where $n_\gamma \approx 400\,{\rm cm}^{-3} \approx
(10^{-10}\, {\rm MeV})^3$ is the photon number density today, so the rate of e.g.
$B' + N_1 \rightarrow \overline{L'} + N_2$ interactions will be much smaller
than the rate of $B' \rightarrow \overline{L'} + N_1 + N_2$ decays,
since $(\Delta m)^3 \gg n_\gamma$. Note that this scenario is only viable
for small enough $N$ masses --- if the mass were large enough that early decoupling
would give too large a relic density, we would need some other mechanism
to reduce the eventual abundance (e.g. annihilation to lighter states).


\section{$B'$-$L'$ mass splitting and $\overline{L'}$ injection velocity}
\label{ApA}

As discussed in Sect.~\ref{sec:splitting}, if the $B' \rightarrow
\overline{L'} + \cdots$ decay has more than two decay products, then we
will obtain a distribution of final velocities for the $\overline{L'}$. In particular, a fraction of them will obtain velocities
higher than some critical value $v_c$ (e.g.\ the escape velocity of a bound
structure). Here, we estimate this fraction, and confirm that for
mass splittings $m_{B'} - m_{L'} \equiv \Delta m$ larger than $m_{L'} v_c$
it is very close to 1, assuming that none of the final or intermediate states
(other than $\overline{L'}$) have masses close to $\Delta m$.

In the $B' \rightarrow \overline{L'} + \cdots$ decay, if the other
decay products carry away 4-momentum $q$, the velocity given to
$\overline{L'}$ corresponds to a Lorentz factor $\gamma$ of
\beq
 \gamma - 1 =
\frac{(\Delta m)^2 - q^2}{2 m_{B'} m_{L'}}~,
\eeq
so if $\Delta m = m_B - m_L$ is small, the velocity is
\beq
v^2 = \frac{(\Delta m)^2 - q^2}{m_{L'}^2} +
O\left(\left(\frac{\Delta m}{m_{L'}}\right)^3\right) + O(v^4)~.
\eeq 
For a two-body decay $B \rightarrow \overline{L} + X$, assuming that $m_X$ is
small compared to $\Delta m$ gives $v \approx \frac{\Delta m}{m_{L'}}$.
For a higher-multiplicity final state, some fraction of the
decays will result in $v < v_c$, i.e.\
those for which $q^2$ is only just below $(\Delta m)^2$.
We can write the differential decay width to a $n$-body final state as
\begin{equation}
{\rm d}\Gamma = \frac{(2 \pi)^4}{2 M} |\mathcal{A}|^2 {\rm d}\Phi_n(P;p_1,\dots,p_n)~,
\end{equation}
where $M$ is the mass of the decaying particle, $\mathcal{A}$ is the amplitude
for that particular decay, and ${\rm d}\Phi_n$ is the differential phase space element
for initial momentum $P$ and final momenta $p_1, \dots, p_n$.
Also, we have
\beq
{\rm d}\Phi_n(P;p_1,\dots,p_n) = (2 \pi)^3 {\rm d} q^2 {\rm d}\Phi_2 (P;p_1,q) {\rm d}\Phi_{n-1}(q;p_2,\dots,p_n)~,
\eeq
splitting the decay into an initial two-body step and then a fragmentation of the second body.
In the rest frame of the decaying particle, the two-body phase space element has
the form ${\rm d}\Phi_2 \propto {\rm d}\Omega \frac{|\vec{p}|}{E}$. 
From above, for small $\Delta m$ we have $|\vec{p}|^2 \approx (\Delta
m)^2 - q^2$, so writing $q^2 = (\Delta m)^2 - \delta q^2$, the leading
order behaviour of ${\rm d}\Phi_2$ with $\delta q^2$ is $\sqrt{\delta
q^2}$. Since the mass dimension of ${\rm d}\Phi_n$ is $2n - 4$, if
the total mass of the other decay particles is small compared to
$\sqrt{\delta q^2}$, then ${\rm d}\Phi_{n-1}$ must vary like $(\delta
q^2)^{n-3}$. So overall, $ {\rm d}\Phi_n$ will vary as $(\delta
q^2)^{n-3/2}$.

The phase space volume with $v < v_c$ corresponds to that with $\delta
q^2$ below a critical value, and we want to compare the total width for
that volume to the total width overall. If we do not have
intermediate states with masses close to $\Delta m$ (heavier states
have a roughly constant effect on $|\mathcal{A}|$, while lighter states
enhance it towards smaller $q^2$), then we can obtain an approximate
upper bound on the ratio of widths by comparing the phase space volumes. From
above, as long as the other final state masses are small compared to
$\Delta m$, this will be well approximated by $\left(\frac{\delta
q^2}{(\Delta m)^2}\right)^{n - 3/2}$. 

To take an example, suppose that
$m_{B'} = 20~\rm{GeV}$, $m_{L'} = 10~{\rm GeV}$, and $v_c = 500~{\rm km}{\rm
s}^{-1}$, and that the decay is $B' \rightarrow \overline{L'} + \phi +
\phi$, where the $\phi$ are massless states. Then, $v < v_c$ corresponds
to $\delta q^2 < 5.6 \times 10^{-2}~{\rm GeV}^2$, which is $1.27
\times 10^{-8}$ of the 3-body phase space, while $\left(\frac{\delta
q^2}{(\Delta m)^2}\right)^{3/2} = 1.31 \times 10^{-8}$.
Since $v^2 \approx \frac{\delta q^2}{m_{L'}^2}$, in general our
volume ratio is $\left(\frac{m v_c}{\Delta m}\right)^{2n - 3}$
(for $\Delta m > m v_c$; otherwise, the whole of the 
phase space volume corresponds to a velocity kick $< v_c$).


\section{Calculating the Galactic $\overline{L'}$ distribution}
\label{ApB}

Suppose that we have a steady-state distribution of (collisionless)
DM particles $B'$ in a gravitational potential $\Phi$. A small fraction
of these then decay, with the decay products including a particle
$\overline{L'}$ of only very slightly smaller mass, whose relative velocity
is consequently non-relativistic. The problem is to calculate the
steady-state distribution of the resulting decay product population.

We could approach this in a brute-force way by sampling from
a large number of $B'$ starting positions and velocities, sampling
from the possible relative $\overline{L'}$ velocities, then calculating the
resulting $\overline{L'}$ orbit and accumulating the time spent at given
$\vec{x}$ and $\vec{v}$ in this orbit into the overall $\vec{x},\vec{v}$
distribution (more sophisticated analyses such as~\cite{PeterHalos}
take some variation of this approach). However, we can simplify the problem slightly by assuming
that the $B'$ distribution, and the gravitational potential, are both
spherically symmetric (and that the $B'$ distribution is non-rotating).
Although this ignores various effects, their impact should be minor~\cite{BT}:
\begin{itemize}
\item {\em The evolution of the galactic potential with time}: most obviously, particles
injected before matter has collapsed into galaxies will not behave as outlined, and may be
captured into galaxies later if their velocity is sufficiently low. 
These will then have the same distribution as the `parent' $B'$ and $L'$
particles.

We can put a rough upper bound on this effect by estimating the proportion of the early-emitted $\overline{L'}$ that
are captured into galaxies.
Free-streaming with a velocity $u$ suppresses perturbations on conformal scales with
$k > H(t)a(t)/u(t)$, as such particles will escape potential wells.
 Since Hubble expansion reduces the velocity of non-relativistic particles
as $1/a$, we have $u(t) = u(t_1) a(t)/a(t_1)$, where $t_1$ is the time of injection,
and $u(t_1)$ is the velocity kick. So, the critical $k$ at the present time is
$k_c = \frac{H_0 a(t_1)}{u(t_1)}$. If $k_c \lesssim 100~{\rm kpc}$, the
scale relevant to galaxies, then the injected particles will never have clustered
into galaxies. Conversely, only states emitted with $a(t_1) \lesssim H_0 u(t_1)^{-1} / (100\,{\rm kpc})$
will cluster. 
During matter domination, $a \sim t^{2/3}$, and since $H_0  / (100\,{\rm kpc})
\simeq 7\,{\rm km}\,{\rm s}^{-1}$, it follows that for an $\overline{L'}$ decay product to be captured it must be produced prior to 
\beq
t_1 \lesssim\left(\frac{7\,{\rm km}\,{\rm s}^{-1}}{u(t_1)}\right)^{3/2}t_0~.
\eeq
Taking example velocity kicks from earlier plots, if $u(t_1) \approx 70\,{\rm km}\,{\rm s}^{-1}$,
around 3\% of the $\overline{L'}$ emitted cluster in this way, 
and taking $u(t_1) \approx 220 \,{\rm km}\,{\rm s}^{-1}$ gives $\sim 5 \times 10^{-3}$
of the $\overline{L'}$ with the `parent' distribution.
From inspection of Fig.~\ref{fig:nfwAlpha}, the modified distribution is always a significantly
larger fraction of the original density than those values. 

\item  {\em The shape of the DM halo}:
$N$-body simulations appear to favour ellipsoidal halos, but with a ratio
of longest/shortest axes around 0.6 rather than more extreme values.
\item  {\em Angular momentum of the DM halo}:
simulations indicate that the velocity bias due to net rotation
is insignificant compared to the velocity dispersion, so should
not give a large effect.
\item  {\em The potential of the galactic disk}: there is a degeneracy between
the contribution of the halo and the disk to the mass
of the inner few parsecs of the galaxy. As a result,
in the cases where the velocity kick has most effect (i.e. DM profiles
with a central density cusp to be smoothed out) the contribution
of the disk is less important.
\end{itemize}

By the Strong Jeans Theorem~\cite{BT}, the steady state phase-space
distribution of a system of collisionless particles moving in a
spherical potential can be expressed as $f = f(\varepsilon, \vec{L})$,
where $\varepsilon$ is the binding energy and $\vec{L}$ is the angular
momentum (both per unit mass). Furthermore, since the $B' \rightarrow
\overline{L'} + \cdots$ decay is spherically symmetric, the distribution of the
$\overline{L'}$ in phase space must be of the form $g = g (\varepsilon, L^2)$.
So, if we start out with a distribution function (DF) $f(\vec{x},
\vec{v}) = f(\varepsilon, L^2)$ for the $B'$, we can derive the
`post-injection' DF 
\beq
h(\vec{x}, \vec{v}) = \int  {\rm d}^3 v' K(|v - v'|) f(\vec{x}, \vec{v}')~,
\eeq
 for the $\overline{L'}$, where $K(\Delta v)$ corresponds to
the probability of injecting with a given velocity change.
In general, $h$ will not be a steady-state distribution, as can be
seen by considering e.g.\ a high-central-density profile with a
cool core, which will be smoothed out by a large velocity boost.
However, since $\varepsilon$ and $L^2$ are preserved along particle
orbits, the number of particles in a volume $ {\rm d} \varepsilon \,  {\rm d}
L^2$ of $(\varepsilon, L^2)$ space will be the same for $h$ and for
the steady-state distribution $g$. Then, since $g$ depends only on
$\varepsilon$ and $L^2$, we can recover it from the $ {\rm d} \varepsilon \,  {\rm d}
L^2$ density $p$, via $g \,  {\rm d}V = p \,  {\rm d}\varepsilon \,  {\rm d}L^2$, where
$ {\rm d}V =  {\rm d}^3 x \,  {\rm d}^3 v$ is phase space volume.
Explicitly,
\begin{align}
\frac{ {\rm d}V}{ {\rm d}\varepsilon  {\rm d}L^2} &= \int  {\rm d}^3 x \int  {\rm d}^3 v \, \delta \left(\varepsilon - \left(\psi - \frac{1}{2}v^2\right)\right) \delta (L^2 - r^2 v_\perp^2) \\
&= 2 \pi \int  {\rm d}^3 x \left(\frac{v}{\sqrt{r^2 v^2 - L^2}}\right)_{v^2 = 2(\psi(r) - \varepsilon)}~,
\end{align}
where $v_\perp$ is the perpendicular-to-radial velocity, and $\psi(r) \equiv -\Phi(r)$ is the maximum binding energy at $r$. Similarly,
\begin{equation}
p(\varepsilon, L^2) = \int  {\rm d}^3 x \int  {\rm d}^3 v \, f(\vec{x}, \vec{v}) \delta \left(\varepsilon - \left(\psi - \frac{1}{2}v^2\right)\right) \delta (L^2 - r^2 v_\perp^2)~,
\end{equation}
and in particular, if the velocity distribution is everywhere isotropic
($h(\vec{x}, \vec{v}) = h(\vec{x}, |v|)$), then
\begin{equation}
p(\varepsilon, L^2) = 
2 \pi \int  {\rm d}^3 x \left(\frac{v h(x,v)}{\sqrt{r^2 v^2 - L^2}}\right)_{v^2 = 2(\psi(r) - \varepsilon)}~.
\end{equation}
From $g(\varepsilon, L^2)$, 
we can find the number density $\rho(r)$ by integrating over the appropriate ranges of $\varepsilon$
and $L^2$,
\begin{equation}
\rho(r) = \int  {\rm d}^3 \vec{v} \, g(r, \vec{v}) = \int_{v^2 < 2 \psi(r)}  {\rm d}^3 \vec{v} \, g(\psi(r) - v^2/2, r^2 v^2 \sin^2\theta)~.
\end{equation}

To find a plausible initial DF for $B'$, we can assume that the velocity
distribution is everywhere isotropic. This implies that the DF depends
only on $\varepsilon$ (it is said to be `ergodic'). Starting from a number
density $\rho(r)$ in a potential $\psi(r)$, there is a unique ergodic DF $f(\varepsilon)$
giving $\rho$~\cite{BT}, described by the Eddington formula
\begin{equation}
f(\varepsilon) = \frac{1}{\sqrt{8}\pi^2}\frac{ {\rm d}}{ {\rm d}\varepsilon}\int_0^\varepsilon \frac{ {\rm d}\psi}{\sqrt{\varepsilon - \psi}} \frac{ {\rm d}\rho}{ {\rm d}\psi}~.
\end{equation}
For the figures in the main text, we also (for ease of implementation) make the approximation 
that the $\overline{L'}$ distribution $g$ is ergodic, i.e.\ that it
does not depend on $L^2$. This can be a poor approximation for large
velocities, as can be seen by considering a cool, high-density centre
subjected to a large velocity kick, as this results in most trajectories
at large distances being approximately radial. However, for
the not-too-large velocity kicks we consider here (as required for 
observable annihilation signals), and for the smaller-radius
regions we are most concerned about, the approximation is acceptable. 
For example, taking the $B$ profile to be the NFW $\gamma=1.2$ one
considered in Fig.~\ref{fig:nfwHooper}, and working out the full
$\overline{L'}$ distribution function $g(\varepsilon, L^2)$ (for $\Delta
v = 109~{\rm km}{\rm s}^{-1}$) at some representative values, we can
estimate that the error in $\rho$ at $r = 10 r_{\rm deg}$ (in the
notation of Fig.~\ref{fig:nfwHooper}) will be at most a few percent.

\end{document}